\documentclass{emulateapj}
\usepackage{apjfonts}
\usepackage{graphicx}

\slugcomment{To appear in ApJ}

\newcommand\ks{\kappa_s}

\newcommand\rv{\mbox{\boldmath $r$}}
\newcommand\xv{\mbox{\boldmath $x$}}
\newcommand\uv{\mbox{\boldmath $u$}}
\newcommand\du{\delta\uv}
\newcommand\av{\mbox{\boldmath $\alpha$}}

\newcommand\avg[1]{\left\langle{#1}\right\rangle}
\newcommand\Rein{R_{\rm Ein}}
\newcommand\Sigcr{\Sigma_{\rm crit}}
\newcommand\tot{{\rm tot}}
\newcommand\mhat{{\hat m}}

\newcommand\refeq[1]{Equation~(\ref{eq:#1})}
\newcommand\refeqs[2]{Equations~(\ref{eq:#1}) and (\ref{eq:#2})}
\newcommand\reffig[1]{Figure~\ref{fig:#1}}
\newcommand\reffigs[2]{Figures~\ref{fig:#1} and \ref{fig:#2}}
\newcommand\reftab[1]{Table~\ref{tab:#1}}
\newcommand\reftabs[2]{Tables~\ref{tab:#1} and \ref{tab:#2}}
\newcommand\refsec[1]{Section \ref{sec:#1}}

\begin{document}

\title{A New Channel for Detecting Dark Matter Substructure in Galaxies: \\
Gravitational Lens Time Delays}

\author{
  Charles R.\ Keeton\altaffilmark{1}
  and
  Leonidas A.\ Moustakas\altaffilmark{2}
}
\affil{
  $^1$
  Department of Physics \& Astronomy, Rutgers University,
  136 Frelinghuysen Road, Piscataway, NJ 08854, USA \\
  $^2$
  Jet Propulsion Laboratory, California Institute of Technology, 4800
  Oak Grove Drive, MS 169-327, Pasadena, CA 91109, USA
}


\begin{abstract}
We show that dark matter substructure in galaxy-scale halos perturbs
the time delays between images in strong gravitational lens systems.
The variance of the effect depends on the subhalo mass function,
scaling as the product of the substructure mass fraction and a
characteristic mass of subhalos (namely
$\langle m^2\rangle/\langle m\rangle$).  Time delay perturbations
therefore complement gravitational lens flux ratio anomalies and
astrometric perturbations by measuring a different moment of the
subhalo mass function.  Unlike flux ratio anomalies, ``time delay
millilensing'' is unaffected by dust extinction or stellar
microlensing in the lens galaxy.  Furthermore, we show that time
delay {\em ratios} are immune to the radial profile degeneracy
that usually plagues lens modeling.  We lay out a mathematical
theory of time delay perturbations and find it to be tractable
and attractive.  We predict that in ``cusp'' lenses with close
triplets of images, substructure may change the arrival-time
order of the images (compared with smooth models).  We discuss
the possibility that this effect has already been observed in
RX J1131$-$1231.
\end{abstract}


\section{Introduction}

The cold dark matter (CDM) paradigm predicts that 5\%--10\% of each
galaxy's mass is bound up in subhalos left over from the hierarchical
formation process.  In the Local Group, the predicted number of dark
matter subhalos significantly exceeds the observed number of dwarf
galaxy satellites \citep[e.g.,][]{moore,klypin,missing,koposov}.
The abundance of substructure depends on competition between
accretion of new subhalos from the environment and destruction
of old subhalos by tidal forces
\citep[e.g.,][]{TB01,TB04,benson,ZB03,koush,oguri-sub,vdb,Z05}.
Also, the number of subhalos that ``light up'' and become visible
as satellite galaxies depends on whether subhalos are able to retain
their gas against photoevaporation, and on the efficiency of galaxy
formation in low-mass systems \citep[e.g.,][]{bullock,somerville,
kravtsov,koposov2,maccio2}.  Measuring the amount of substructure
in galaxy halos, and how it varies with galaxy mass, environment,
and redshift, therefore provides unique access to the astrophysics
of galaxy formation on small scales.

We still know very little about the physical properties of the dark
matter particle, but a number of specific models have been proposed:
dark matter could be sterile neutrinos, or supersymmetric particles,
or a manifestation of extra dimensions, or even a product from the
decay of any of these particles
\citep[e.g.,][]{dodelson,cheng,feng,strigari-decay}.  All of those
possibilities are compatible with observations that probe the universe
on large scales. However, they make some different predictions about
the amount of dark matter substructure; for example, any type of
``warm'' dark matter can lead to a suppression of power on small
scales \citep[e.g.,][]{WDM,SIDM,ZB03}.  Studying galaxy substructure
provides the opportunity to test such models and obtain important
astrophysical evidence about the fundamental nature of dark matter.

Strong gravitational lensing is a simple geometric phenomenon that
gives us a valuable tool for studying mass in distant galaxies,
including substructure.  Lensing effects are succinctly encoded in
the properties of the time delay surface.  Consider light emitted
by a background source at angular position $\uv$ from the center
of a foreground lens with (projected) potential $\phi$.\footnote{The
lens potential $\phi$ is a scaled version of the two-dimensional
gravitational
potential.  Specifically, it satisfies the Poisson equation
$\nabla^2\phi = 2 \Sigma/\Sigcr$ where $\Sigma$ is the surface mass
density of the lens, and $\Sigcr$ is the critical surface density
for lensing.}  If the light reaches us from the angular image
position $\xv$, the excess travel time relative to a hypothetical
light ray that travels directly from the source with no deflection
is
\begin{equation} \label{eq:tdel}
  \tau(\xv) = t_0 \left[ \frac{1}{2} \left|\xv-\uv\right|^2
    - \phi(\xv) \right] ,
  \quad
  t_0 = \frac{1+z_l}{c}\ \frac{D_l D_s}{D_{ls}}\ .
\end{equation}
By Fermat's principle, images form at stationary points of the time
delay surface, i.e., positions $\xv$ such that $\nabla\tau(\xv) = 0$.
(This directly yields the lens equation, usually written as
$\uv = \xv - \nabla\phi$.)  The magnifications of the images are
determined by second derivatives of $\tau$ (and hence depend on
second derivatives of the lens potential $\phi$).  What is usually
measured is the differential time delay between two images,
$\Delta t_{ij} = \tau(\xv_j) - \tau(\xv_i)$.  (See the review of
strong lensing by \citealt{saasfee} for more details.)

Strong lensing provides the only way to detect substructure directly
(i.e., by virtue of its gravity) in galaxies outside the Local Group.
Small mass clumps in the lens galaxy can strongly perturb lensed
images.  The spatial perturbations are determined by first derivatives
of the lens potential: they have angular scales of milli-arcseconds
for ``millilensing'' by dark matter subhalos
\citep[e.g.,][]{MS,MM,chiba,DK,metcalf2237,chenastro}, or
micro-arcseconds for ``microlensing'' by stars in the lens galaxy
\citep[e.g.,][]{WMS95,wyithe,SW02,csk-micro,congdon-micro}.
Since lensing magnifications depend on second derivatives of the lens
potential, they are even more sensitive to substructure: magnification
perturbations can be of order unity and are therefore very apparent,
especially in 4-image lenses \citep[e.g.,][]{MZ02,cuspreln,foldreln}.

At optical and X-ray wavelengths quasar emission regions are small
enough that lens flux ratios are sensitive to both dark matter subhalos
and stars \citep[e.g.,][]{csk-micro,0924,blackburne,pooley}; this
makes it difficult to isolate millilensing and study dark matter
substructure.  By contrast, at radio wavelengths the quasar source
is thought to be large enough to smooth over the effects of stars and
be insensitive to microlensing \citep[but see][]{1600micro}.  The
flux ratios of radio lenses (especially 4-image, or quad, lenses)
have therefore been the tool of choice for studying millilensing.
The amount of substructure needed to explain radio lens flux ratios
is broadly consistent with CDM predictions \citep{DK}.  The greatest
limitation of this method is the small number of radio quads known
currently \citep[e.g.,][]{browne}.

\begin{deluxetable}{ccrrrc}
\tablewidth{0pt}
\tablecaption{Reference images: ``Fold'' lens}
\tablehead{
  \colhead{Image} &
  \colhead{New Label} &
  \colhead{$x\ (\arcsec)$} &
  \colhead{$y\ (\arcsec)$} &
  \colhead{$\mu$} &
  \colhead{$\Delta t$ (days)}
}
\startdata
 C     & M1 & $ 0.343$ & $ 1.360$ & $  3.85$ & $\equiv 0$ \\
 A1 & M2 & $-0.948$ & $-0.697$ & $ 13.66$ & $10.77$ \\
 A2 & S1 & $-1.098$ & $-0.206$ & $-12.50$ & $10.93$ \\
 B     & S2 & $ 0.700$ & $-0.652$ & $ -3.02$ & $17.82$
\enddata
\tablecomments{
Reference smooth images for our sample fold lens similar to PG 1115+080.
All coordinates are given in arcseconds with respect to the center of
the lens galaxy at $(0,0)$.  The time delays are given with respect to
the leading image.  The sign of the magnification ($\mu$) represents
the image parity.  Col.\ 1 gives the traditional image names for
PG 1115+080, while Col.\ 2 gives our new labels: ``M'' and ``S''
indicate (respectively) an image at a local minimum or saddlepoint
of the time delay surface, while ``1'' and ``2'' indicate
(respectively) the leading and trailing image of each type.
}\label{tab:foldimg}
\end{deluxetable}

In this paper we show that lens {\em time delays} provide an
exciting new way to probe dark matter substructure, with several
distinct advantages.  As we shall see, time delays are not affected
by microlensing, so we can use optical as well as radio data to
probe dark matter substructure; this is important because there
will be thousands of lenses discovered in new optical surveys
\citep[e.g.,][]{LSSTlens,kuhlen,SNAPlens,csk-find}, complementing
new samples of radio lenses \citep[e.g.,][]{SKAlens}.  Time delays
are sensitive to the mass function of dark matter subhalos in a way
that flux ratios are not; so they offer a good opportunity to probe
the masses of subhalos in distant galaxies.  By all indications
the theory of time delay millilensing is very tractable; having
a formal theory will provide a rigorous foundation for substructure
studies, and may even allow us to do some of the statistics
analytically. Appropriate time delay measurements are feasible
now, and truly revolutionary datasets will become available in
the foreseeable future.

For the purpose of concreteness, we assume a cosmology with
$\Omega_M = 0.3$, $\Omega_\Lambda = 0.7$, and
$H_0 = 70$ km s$^{-1}$ Mpc$^{-1}$, and we adopt specific values
for the lens and source redshifts (motivated by particular lens
systems, as discussed below).  Modifying the cosmology or the lens
and source redshifts would simply rescale the time delays through
the lensing time scale $t_0$.

\begin{deluxetable}{ccrrrc}
\tablewidth{0pt}
\tablecaption{Reference images: ``Cusp'' lens}
\tablehead{
  \colhead{Image} &
  \colhead{New Label} &
  \colhead{$x\ (\arcsec)$} &
  \colhead{$y\ (\arcsec)$} &
  \colhead{$\mu$} &
  \colhead{$\Delta t$ (days)}
}
\startdata
 C & M1 & $-1.717$ & $-1.697$ & $ 10.76$ & $\equiv 0$ \\
 B & M2 & $-2.334$ & $ 0.612$ & $ 11.92$ & $  0.25$ \\
 A & S1 & $-2.305$ & $-0.577$ & $-19.68$ & $  1.22$ \\
 D & S2 & $ 0.796$ & $ 0.315$ & $ -0.99$ & $120.08$
\enddata
\tablecomments{
Similar to \reftab{foldimg}, but for our sample cusp lens similar to
RX J1131$-$1231.  We again introduce new image labels such that M1 is
the leading minimum, M2 is the trailing minimum, S1 is the leading
saddle, and S2 is the trailing saddle.
}\label{tab:cuspimg}
\end{deluxetable}

\section{Methods}

\subsection{Reference smooth lens}
\label{sec:smooth}

We examine the effects of substructure by comparing the properties of
images produced by a galaxy with substructure to those produced by an
equivalent smooth galaxy.  We model the smooth galaxy using a
pseudo-Jaffe profile with scaled surface mass density
\begin{equation}
  \kappa_\tot(r) = \frac{\Sigma(r)}{\Sigcr}
  = \frac{b_\tot}{2} \left( \frac{1}{r}
    - \frac{1}{\sqrt{a^2+r^2}} \right) ,
\end{equation}
where $\Sigcr = (c^2 D_s)/(4\pi G D_l D_{ls})$ is the critical surface
density for lensing, where $D_l$, $D_s$, and $D_{ls}$ are angular
diameter distances between the observer and lens, the observer and
source, and the lens and source, respectively.  We have written the
equation for a circular lens, but it is straightforward to obtain
an elliptical model by replacing $r$ with an appropriate elliptical
radius.  The pseudo-Jaffe model is equivalent to the singular
isothermal model for $r \ll a$, but the density falls more quickly
at large radii $r \gg a$ to keep the total mass finite,
$M_\tot = \pi a b \Sigcr$.  The scale radius $a$ can therefore be
thought of as a sort of truncation radius, although the truncation
is not sharp.  We set $a$ to be 300 kpc, although the particular
choice has little effect on our results provided it is well outside
the Einstein radius (see \refsec{calc} for more discussion).  The
parameter $b_\tot$ is the Einstein radius of the lens.

We calibrate the pseudo-Jaffe model by fitting it to observed lens
systems.  The fit is not quantitatively precise because we model the
lens environment using a simple external tidal shear, when in truth
the lens may lie in a group of galaxies that has a more complicated
effect \citep[e.g.,][]{kundic97,iva}.  For the present work, we seek
only to obtain reasonable smooth mass models, not to replicate any
observed lens in rich detail.  

\begin{figure*}[t]
\centerline{
  \includegraphics[width=2.0in]{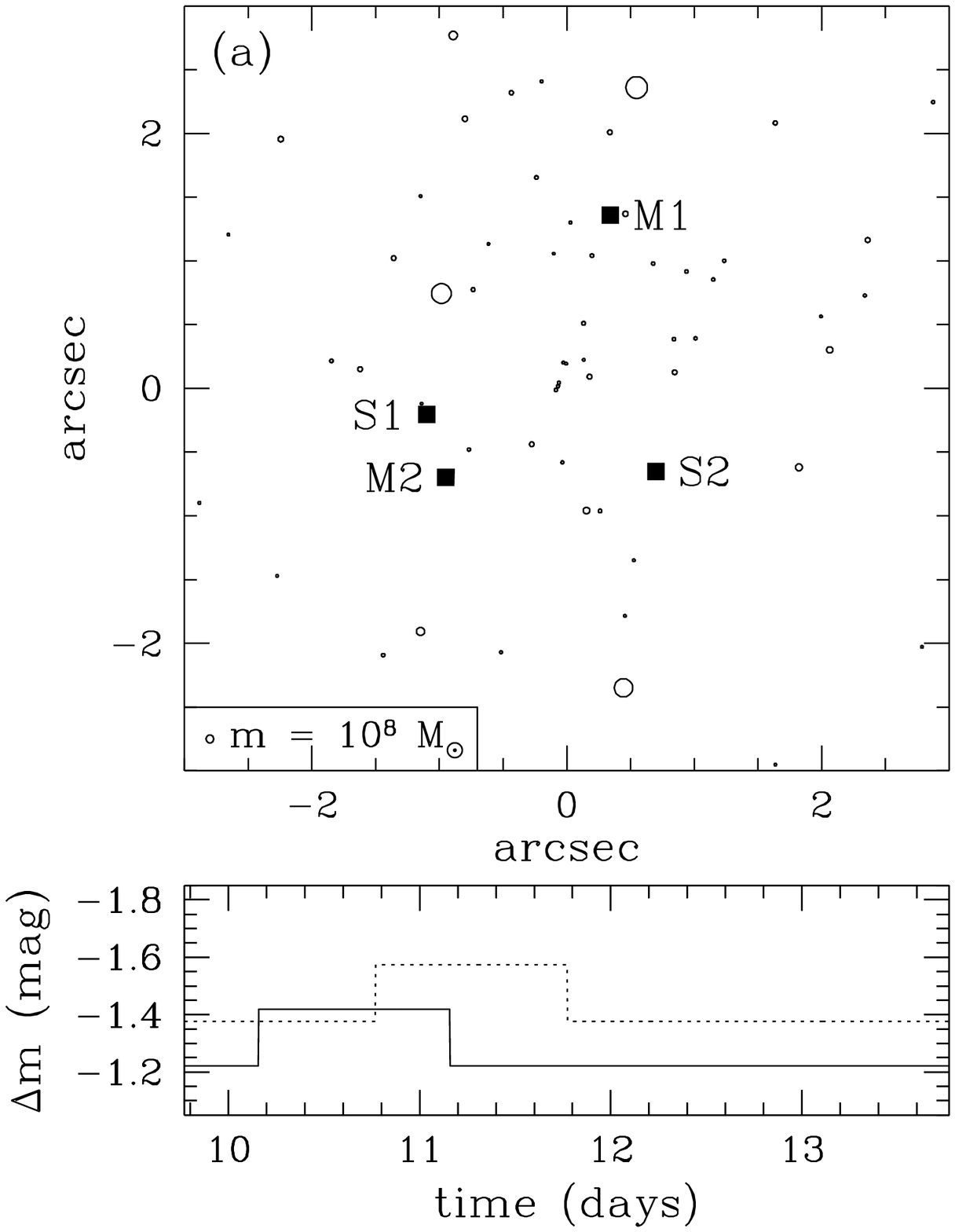}
  \includegraphics[width=2.0in]{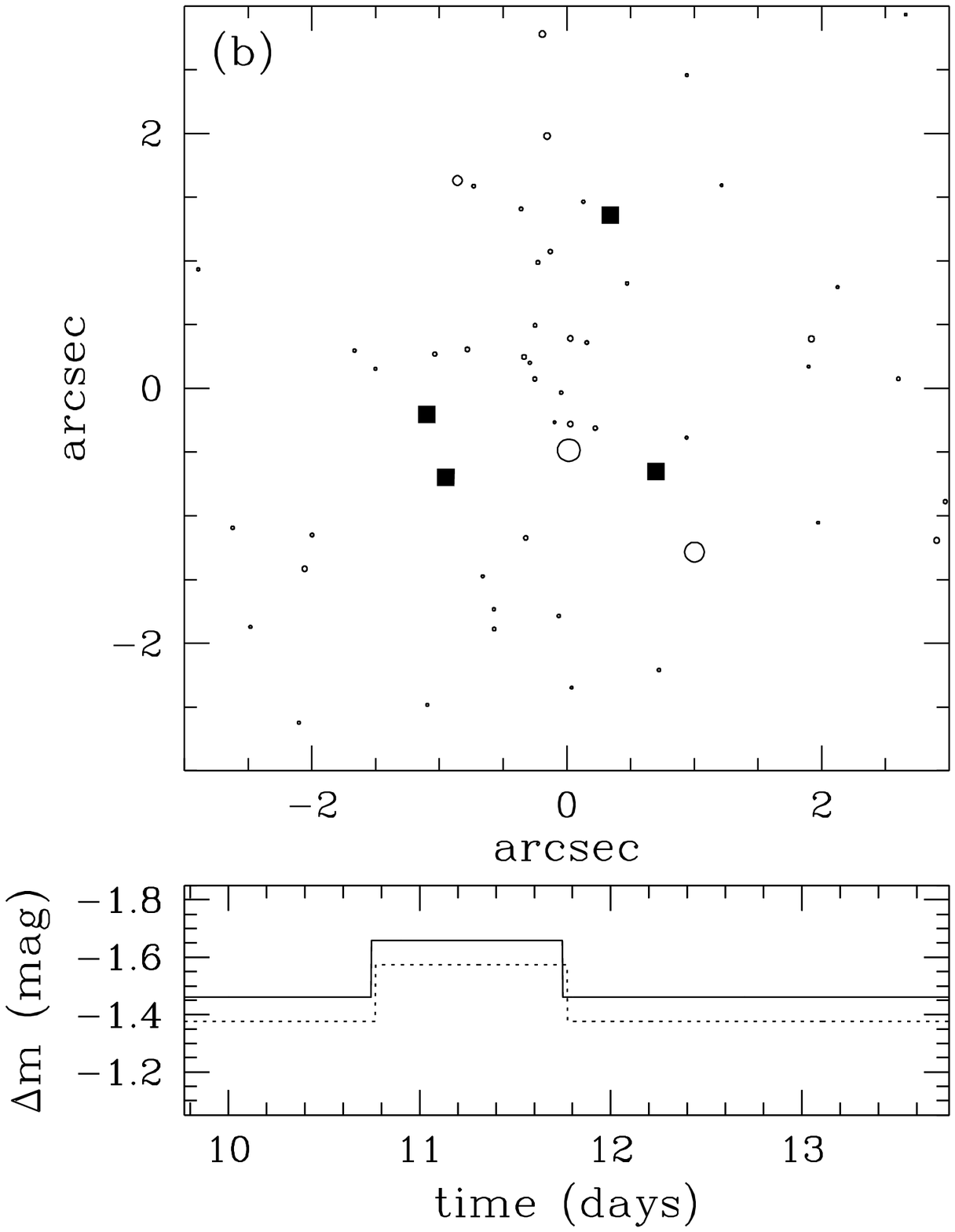}
  \includegraphics[width=2.0in]{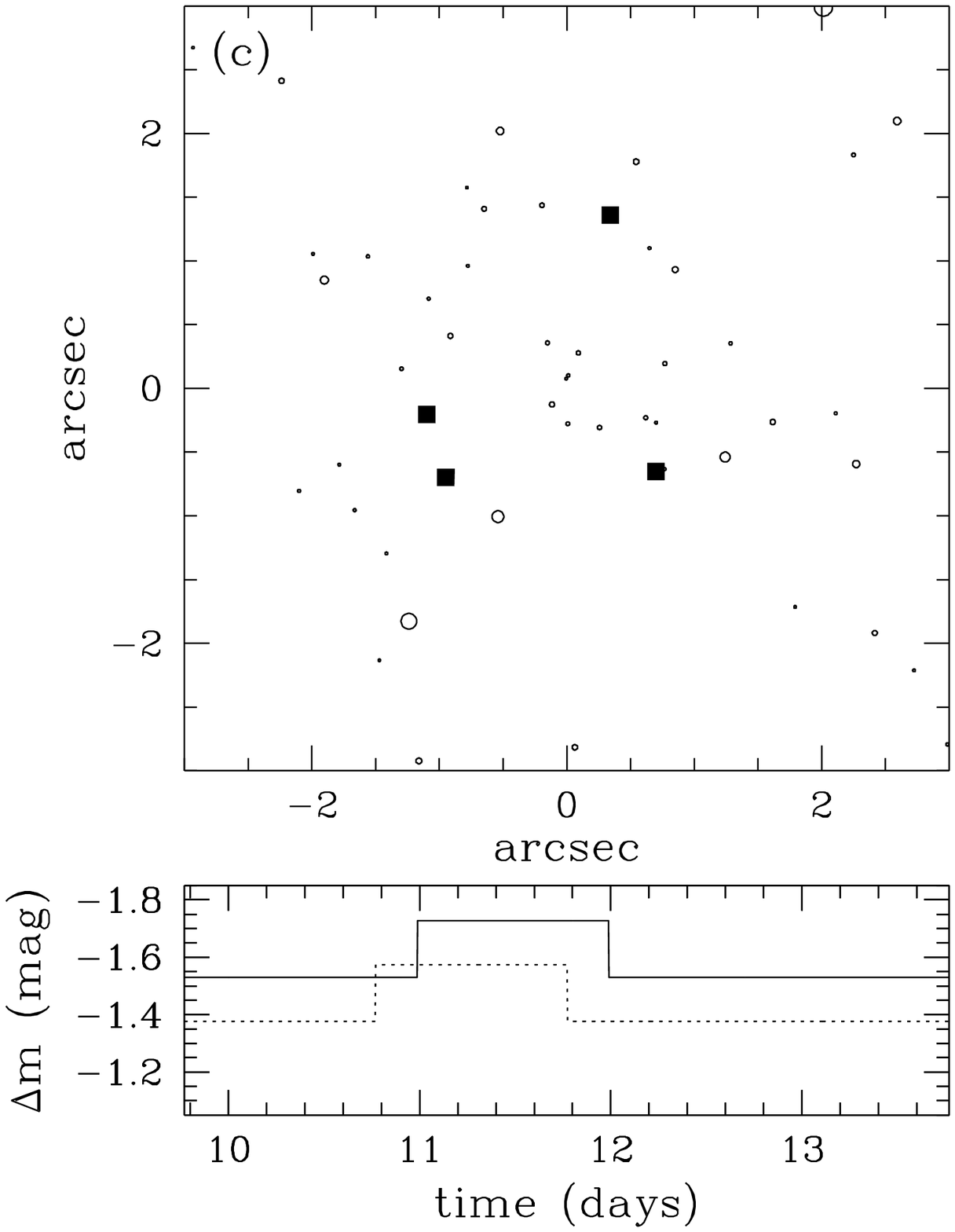}
}
\caption{
Upper panels:
the squares show the images for our sample fold configuration
(compare with \reftab{foldimg}).  In terms of the traditional image labeling
for PG 1115+080, we have M1=C, M2=A$_1$, S1=A$_2$, and S2=B.  Circles
show the Einstein radii of mass clumps; panel (a) includes a key with
$m=10^8\,M_\odot$ for reference.  The three panels show random
realizations of a substructure population in which the substructure
mass fraction is $f_s = 0.01$, and the subhalo mass function is
$dN/dm \propto m^{-1.8}$ over the range $10^7$--$10^9\,M_\odot$.
Lower panels:
sample light curves for image M2.  In these examples, we assume
image M1 brightens by 20\% at $t=0$ and returns to its original
flux at $t=1$ day.  We plot the magnitude difference
$\Delta m = m_{M2} - m_{M1}$.  In each panel the dotted curves shows
the light curve for the reference smooth model, while the solid
curve shows the actual light curve for the specific realization of
substructure.  Notice that the substructure perturbs both the flux
ratio and time delay of image M2 relative to image M1.  (The results
for other images are similar.)
}\label{fig:foldimg}
\end{figure*}

We create a lens with a ``fold'' image configuration similar to that
of PG 1115+080 \citep{weymann}.  The lens galaxy redshift is
$z_l = 0.31$ and the source redshift is $z_s = 1.72$, so the lensing
time scale is $t_0 = 46.5$ days (see \refeq{tdel}).  Our reference
smooth model is circular with Einstein radius $b_\tot = 1\farcs 16$,
and it has external shear $\gamma = 0.12$ at position angle
$\theta_\gamma = 65^\circ$ (east of north).  We place a source at
position $(-0\farcs032, 0\farcs118)$ relative to the center of the
galaxy, and obtain the four images listed in \reftab{foldimg} and
shown in \reffig{foldimg} below.

We also create a lens with a ``cusp'' image configuration similar to
that of RX J1131$-$1231 \citep{sluse}.  The lens redshift is
$z_l = 0.295$ and the source redshift is $z_s = 0.658$, yielding a
lensing time scale $t_0 = 45.5$ days.  Our reference smooth model
has Einstein radius $b_\tot = 1\farcs85$, ellipticity $e = 0.16$ at
position angle $\theta_e = -56^\circ$ (east of north),
and external shear $\gamma = 0.12$ at position angle
$\theta_\gamma = -83^\circ$.  We place a source at position
$(-0\farcs549, -0\farcs142)$ relative to the center of the galaxy,
and obtain the four images listed in \reftab{cuspimg} and shown in
\reffig{cuspimg}.  While there are quantitative differences
in the time delays between the fold and cusp lens examples, many
of the scalings and conceptual conclusions remain the same.
Therefore in most of the presentation we focus on the fold lens
for clarity.  The exceptions are discussed in \refsec{ordering}.

We should comment on our labeling of the lensed images.  Images
are traditionally labeled by letters (e.g., A, B, C, D) using
criteria chosen by the discoverers; different criteria have been
used, with the result that existing labels are not very informative.
\citet{SW-tdel} instead advocate labeling the images 1--2--3--4 in
arrival-time order, arguing that this labeling is robust (at least
for smooth mass distributions; see \refsec{ordering}), and presenting
rules for using the image configuration to determine the labels.
We assert that it is useful to have labels convey not only the
time ordering but also the type of image as well.  We therefore
use ``M'' or ``S'' to indicate (respectively) an image at a local
minimum or saddlepoint of the time delay surface, combined with
``1'' or ``2'' to indicate (respectively) the leading or trailing
image of each type.  Thus, the four images arranged in arrival-time
order are M1, M2, S1, and S2.  In \reftabs{foldimg}{cuspimg} we
give the correspondence between our labeling scheme and the
traditional letter labels for PG 1115+080 and RX J1131$-$1231.

\subsection{Subhalo population}
\label{sec:popn}

While detailed semi-analytic substructure models are now available
\citep[e.g.,][]{TB01,TB04,benson,ZB03,koush,oguri-sub,vdb,Z05},
for pedagogical purposes it is attractive to adopt simple but
reasonable assumptions about the mass function and spatial
distribution of subhalos
\citep[e.g.,][]{MM,chiba,DK,metcalf2237}.\footnote{A few
millilensing studies have worked directly with $N$-body simulations,
but both the original simulations and the lensing calculations are
computationally expensive \citep{bradac1,bradac2,amara,maccio}.} 
We assume that subhalos trace the total mass distribution, so the
average surface mass density in subhalos at radius $r$ is
$\ks(r) = f_s\,\kappa_\tot(r)$, where $f_s$ is the substructure
mass fraction.

CDM simulations predict that the subhalo mass function is
approximately a power law, $dN/dm \propto m^\beta$ with
$\beta \approx -1.8$ \citep[e.g.,][]{Ghigna00,Helmi02,Gao04,Diemand07}.
In order to explore how subhalo masses affect time delays, we
consider masses in some finite range $m_1 \le m \le m_2$.  The mean
subhalo mass is
\begin{equation}
  \avg{m} = \frac{1+\beta}{2+\beta}\ 
    \frac{m_2^{2+\beta}-m_1^{2+\beta}}{m_2^{1+\beta}-m_1^{1+\beta}}\ .
\end{equation}
Equivalently, we may write the lower mass limit in terms of the mean
mass and dynamic range $q = m_2/m_1$ as
\begin{equation}
  m_1 = \avg{m}\ \frac{2+\beta}{1+\beta}\ 
    \frac{q^{1+\beta}-1}{q^{2+\beta}-1}\ .
\end{equation}
Another quantity that will prove to be useful is the mean squared
mass,
\begin{equation} \label{eq:msq}
  \avg{m^2} = \avg{m}^2\ 
    \frac{q^{1+\beta}-1}{1+\beta}\ 
    \frac{q^{3+\beta}-1}{3+\beta}
    \left(\frac{2+\beta}{q^{2+\beta}-1}\right)^2 .
\end{equation}
When $q=1$ the mass function reduces to a Dirac $\delta$-function
and all subhalos have the same mass.

\reffig{foldimg} shows three realizations of a subhalo population
in which the substructure mass fraction is $f_s = 0.01$ and the
subhalos have masses between $m_1 = 10^7$ and $m_2 = 10^9\,M_\odot$.
For reference, this mass function has
$\avg{m} = 6.2 \times 10^7\,M_\odot$ and
$\avg{m^2}/\avg{m} = 2.8 \times 10^8\,M_\odot$.

\subsection{Subhalo models}
\label{sec:models}

For a subhalo of mass $m$, it is useful to define a scaled mass
\begin{equation} \label{eq:mhat}
  \mhat \equiv \frac{m}{\Sigcr} = \pi \Rein^2\,,
\end{equation}
which has dimensions of area.  Here $\Rein$ is the Einstein radius
of the subhalo if it is a point mass.  In most of our analysis we
do treat subhalos as point masses for simplicity.  The lens potential
of a subhalo is then
\begin{equation}
  \phi = \frac{\mhat}{\pi}\,\ln r\,,
\end{equation}
while the deflection angle is
\begin{equation}
  \alpha = \frac{d\phi}{dr} = \frac{\mhat}{\pi r}\ .
\end{equation}
We also consider subhalos modeled as truncated isothermal spheres.
The truncation occurs in the three-dimensional density profile, so
we have
$\rho \propto r^{-2}$ out to the truncation radius $r_t$, and
$\rho = 0$ for $r > r_t$.  The lensing deflection angle works
out to be
\begin{eqnarray}
  r<r_t:\ \alpha &=& \frac{\mhat}{\pi r} \left[ 1
    - \left(1-\frac{r^2}{r_t^2}\right)^{1/2}
    + \frac{r}{r_t} \cos^{-1} \left(\frac{r}{r_t}\right) \right] ,
    \nonumber\\
  r>r_t:\ \alpha &=& \frac{\mhat}{\pi r}\ .
\end{eqnarray}
It is straightforward to obtain the potential by integrating,
$\phi = \int \alpha\,dr$, but it is not instructive to write the
(somewhat long) expression here.

\subsection{Calculations}
\label{sec:calc}

To obtain a mass model with substructure, we leave a fraction
$(1-f_s)$ of the dark matter mass in a smooth component, and replace a
fraction $f_s$ with subhalos.  The expected number of subhalos is
\begin{equation}
  \avg{N} = \frac{f_s M_\tot}{\avg{m}}
  = \frac{\pi a b_\tot f_s}{\avg{\mhat}}\ ,
\end{equation}
and we draw the actual number from a Poisson distribution with
this mean.  The subhalo positions are assigned randomly as follows.
For the pseudo-Jaffe model the cumulative probability distribution
for the radius is
\begin{equation} \label{eq:Pofr}
  P_r(r) = \frac{\int_{0}^{r} r'\,\ks(r')\,dr'}
    {\int_{0}^{\infty} r'\,\ks(r')\,dr'}
  = 1 + \frac{r}{a} - \left(1+\frac{r^2}{a^2}\right)^{1/2} .
\end{equation}
We pick a random value for $P_r$ uniformly between 0 and 1, and
then invert \refeq{Pofr} to find the random radius.  The explicit
inversion is
\begin{equation}
  r = a\ \frac{P_r(2-P_r)}{2(1-P_r)}\ .
\end{equation}
(Again, these equations are for a circular lens, but the extension
to an elliptical mass distribution is straightforward.)  We then
pick a random azimuthal angle uniformly between 0 and $2\pi$.

The pseudo-Jaffe model formally extends to infinity, but we do not
actually need to consider subhalos at large radii because they have
little effect on the time delays.  In the Appendix, \refeq{Rthresh}
allows us to calculate the rms error that we make on time delays if
we neglect subhalos beyond some radius $R_0$.  We use this equation
to set the threshold radius to ensure that our time delay errors
are less than 0.001 days.  The fact that time delays are not very
sensitive to distant subhalos is the reason that our results do not
depend strongly on the choice of the pseudo-Jaffe ``truncation''
radius.

The subhalo masses are assigned randomly from the power law mass
function.  The cumulative probability distribution for the mass is
\begin{equation} \label{eq:Pofm}
   P_m(m) = \frac{m^{1+\beta}-m_1^{1+\beta}}{m_2^{1+\beta}-m_1^{1+\beta}}\ .
\end{equation}
We pick a random value for $P_m$ uniformly between 0 and 1, and
invert \refeq{Pofm} to find the random mass.  (When the mass
function is a $\delta$-function, all halos have the same mass and
this process is unnecessary.)

The number of subhalos in a simulation varies between a few hundred
and a few hundred thousand, depending on the subhalo masses and the
substructure mass fraction.  We use a tree algorithm \citep{tree} to
compute the net lensing potential, deflection, and magnification
quickly and efficiently.  All of this analysis is included in an
updated version of the public software package {\em gravlens}
\citep{gravlens}.

To find the images as they are affected by substructure, we note
that the shifts in the image positions are small
\citep[e.g.,][]{chenastro}, so it is efficient to start from the
smooth images and perturb the positions iteratively:
\begin{equation}
  \xv^{(i+1)} = \xv^{(i)} + \mu^{(i)} \cdot \du^{(i)}\,,
  \quad \du^{(i)} = \xv^{(i)} - \av^{(i)} - \uv\,,
\end{equation}
where $\av^{(i)}$ and $\mu^{(i)}$ are the deflection vector and
magnification tensor, respectively, evaluated at the current position
$\xv^{(i)}$.  (Here $i$ labels the iteration step, not the image
index.)  Note that $\xv^{(i)} - \av^{(i)}$ represents the source
position associated with the current image position, so $\du^{(i)}$
is the offset in the source plane.  When the image position correctly
solves the lens equation, $\du^{(i)}$ vanishes and the iteration
process converges.

A massive clump near one of the ``macro-images'' can in principle
split it into several ``milli-images''.\footnote{With point mass
clumps, or indeed any clump with a cuspy central density steeper
than $\Sigma \propto r^{-1}$, there is formally a ``micro-image''
very close to each clump.  However, these images are highly
demagnified and hence unimportant.  They are absent when the clump
central density is shallower than $\Sigma \propto r^{-1}$.}  We
check for this possibility by comparing the parity of the recovered
image against the parity of the reference image.  If we start with
a minimum but recover a saddle (or vice versa), we know there are
extra images and we have found one.  Note that if we start with a
minimum and recover a minimum (or likewise for a saddle), that does
not prove there are no extra images.  However, we use the
minimum/saddle cases to estimate the fraction of simulations that
yield extra images, and find it to be small.  We throw away the
small number of simulations that yield extra images.

Once we have found the image positions, we use \refeq{tdel} to
compute the differential time delay between image pairs,
$\Delta t_{ij} = \tau(\xv_j) - \tau(\xv_i)$.

\section{Results}

\subsection{Time delay perturbations}
\label{sec:pert}

\begin{figure}[t]
\epsscale{1.0}
\plotone{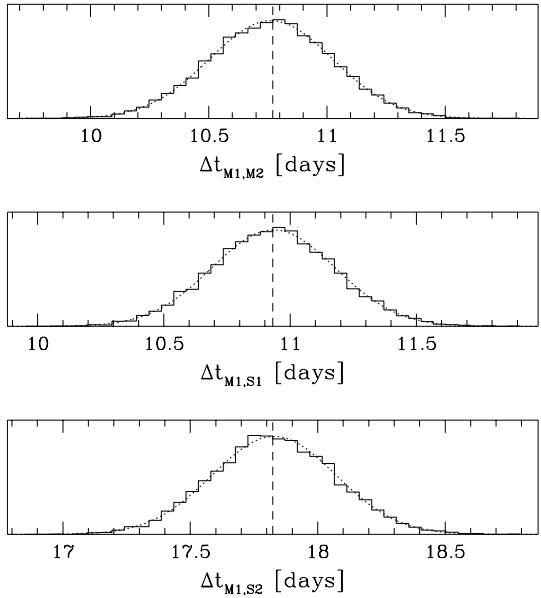}
\caption{
Histograms of the time delays between the images, for $10^4$ Monte
Carlo simulations with substructure mass fraction $f_s = 0.01$ and
subhalo mass $m = 10^8\,M_\odot$.  (The vertical axis scale is
arbitrary.)  The vertical lines indicate the time delays for the
reference smooth model.  Superposed on each histogram is a Gaussian
(dotted line) with the same mean and variance as the simulated data.
}\label{fig:hist}
\end{figure}

We begin by examining our sample fold lens in the presence of
substructure, as shown in \reffig{foldimg}.  In the bottom row of
that figure we show sample light curves for image M2 (assuming for
pedagogical purposes a simple square wave variation in the source);
the results for the other images are similar.  The substructure
clearly changes both the flux ratio and time delay between the
images.  In these examples, the flux ratio is perturbed by
0.1--0.2 mag and the time delay is perturbed by some fraction
of a day.  \citet{oguriH0} also showed (in less detail) that
substructure can affect lens time delays, but he focused on
this as a source of noise in lensing measurements of the Hubble
constant.  We instead propose that time delay perturbations
provide a new way to detect and study substructure itself, if
lens time delays can be measured well enough.

\begin{figure}[t]
\epsscale{1.0}
\plotone{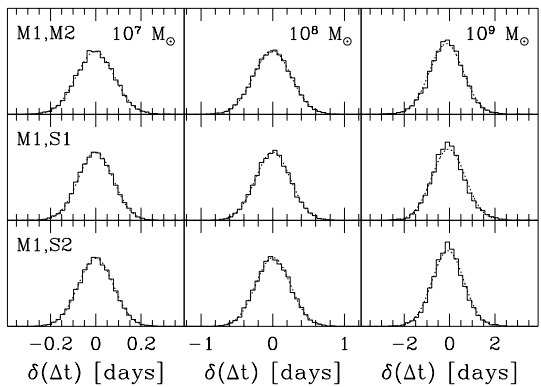}
\caption{
Histograms of the time delays between different images (rows), for
different values of the subhalo mass (columns).  We now plot
$\delta(\Delta t)$, the difference between the time delay with
substructure and the delay for the reference smooth model.  All
cases have $10^4$ Monte Carlo simulations with a substructure mass
fraction $f_s = 0.01$.  Notice that the horizontal scale varies from
one column to the next.  (The vertical axis scale varies as well,
but the units are arbitrary anyway.)  Equivalent Gaussians are again
shown with dotted lines.
}\label{fig:hist-m}
\end{figure}

The remainder of the paper is devoted to exploring how time delay
perturbations depend on the subhalo population.  To that end, we
start with a simple case in which all the subhalos are point masses
with $m = 10^8\,M_\odot$, and the substructure mass fraction is
$f_s = 0.01$.  For each substructure realization we compute the
time delays between the images; we then repeat the process $10^4$
times and make histograms of the time delays, as shown in
\reffig{hist}.  For each image pair, the average time delay
matches the prediction of the reference smooth model very well,
which makes sense because the smooth model can be thought of as
an average over the subhalo population.  These three time delay
histograms appear to be Gaussian.  (The histogram of the time
delay between the close images M2 and S1 is somewhat different,
as discussed in \refsec{ordering}.)  We attribute the Gaussianity
to the Central Limit Theorem: the lens potential is a sum of many
random terms, and such a sum tends toward a Gaussian distribution
if the mean and variance of each term are finite, which we verify
in the Appendix.\footnote{The Central Limit Theorem has not been
used in millilensing before, because it does not obviously apply
to flux ratios.  See the Appendix for more discussion.}  The
standard deviation is $\sigma_t = 0.24$--0.26 days for all three
image pairs shown in \reffig{hist}.

\subsection{Subhalo mass and substructure mass fraction}
\label{sec:m-f}

\begin{figure}[t]
\epsscale{1.0}
\plotone{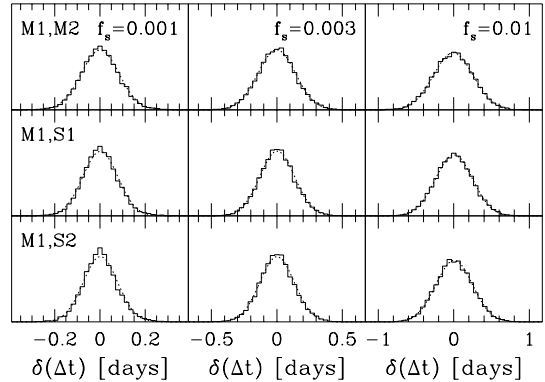}
\caption{
Similar to \reffig{hist-m}, but for different values of the
substructure mass fraction.  All simulations have subhalos with
mass $m = 10^{8}\,M_\odot$.
}\label{fig:hist-f}
\end{figure}

Next we consider the effects of changing the subhalo mass
(\reffig{hist-m}) and the substructure mass fraction
(\reffig{hist-f}).  In order to place different image pairs
on a common scale, we plot histograms of the time delay perturbation
$\delta(\Delta t)$, defined to be the difference between the time
delay with substructure and the time delay for the reference smooth
model.  
All of the histograms appear to be nicely Gaussian, to be centered on
zero, and to have widths that increase with both the subhalo mass and
the substructure mass fraction.

\begin{figure}[t]
\epsscale{1.0}
\plotone{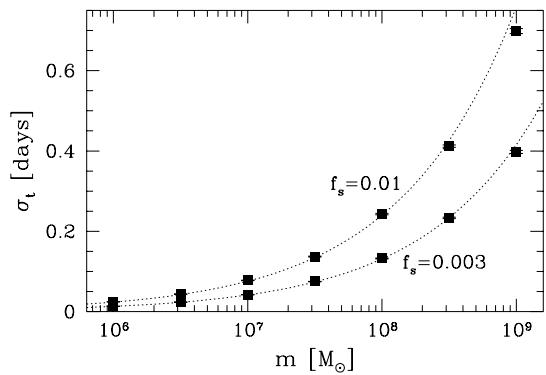}
\caption{
Time delay scatter as a function of subhalo mass.  The points show
results from Monte Carlo simulations (compare \reffigs{hist-m}{hist-f}),
with statistical errorbars
from bootstrap resampling.  The dotted curves show the scaling
$\sigma_t \propto (f_s\,m)^{1/2}$ from \refeq{sigf2} in the
Appendix (after we fit for the proportionality constant using the
$f_s = 0.01$ points).
}\label{fig:mscaling}
\end{figure}

\begin{figure*}[t]
\epsscale{1.0}
\centerline{
  \includegraphics[width=2.5in]{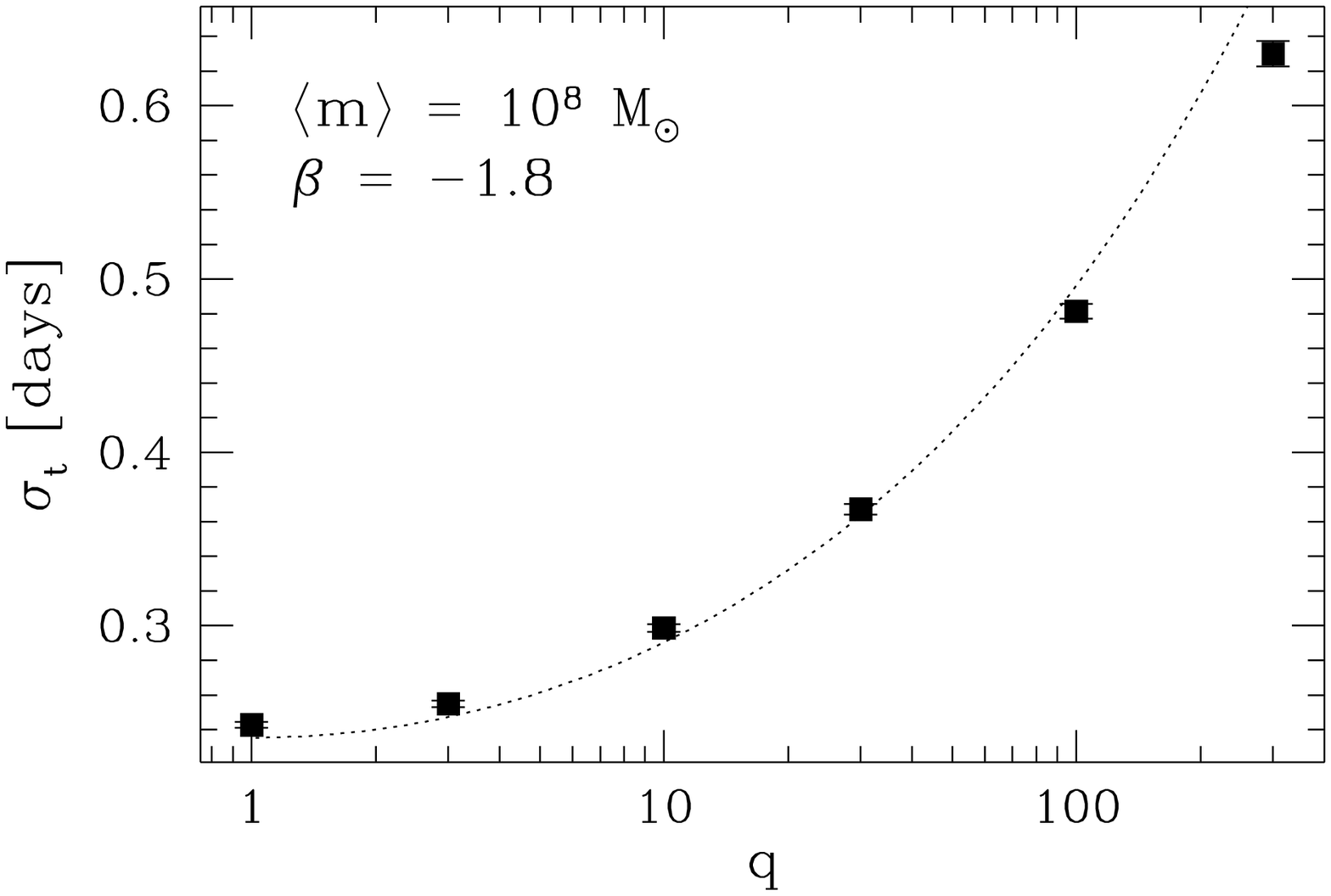}
  \hspace{0.15in}
  \includegraphics[width=2.5in]{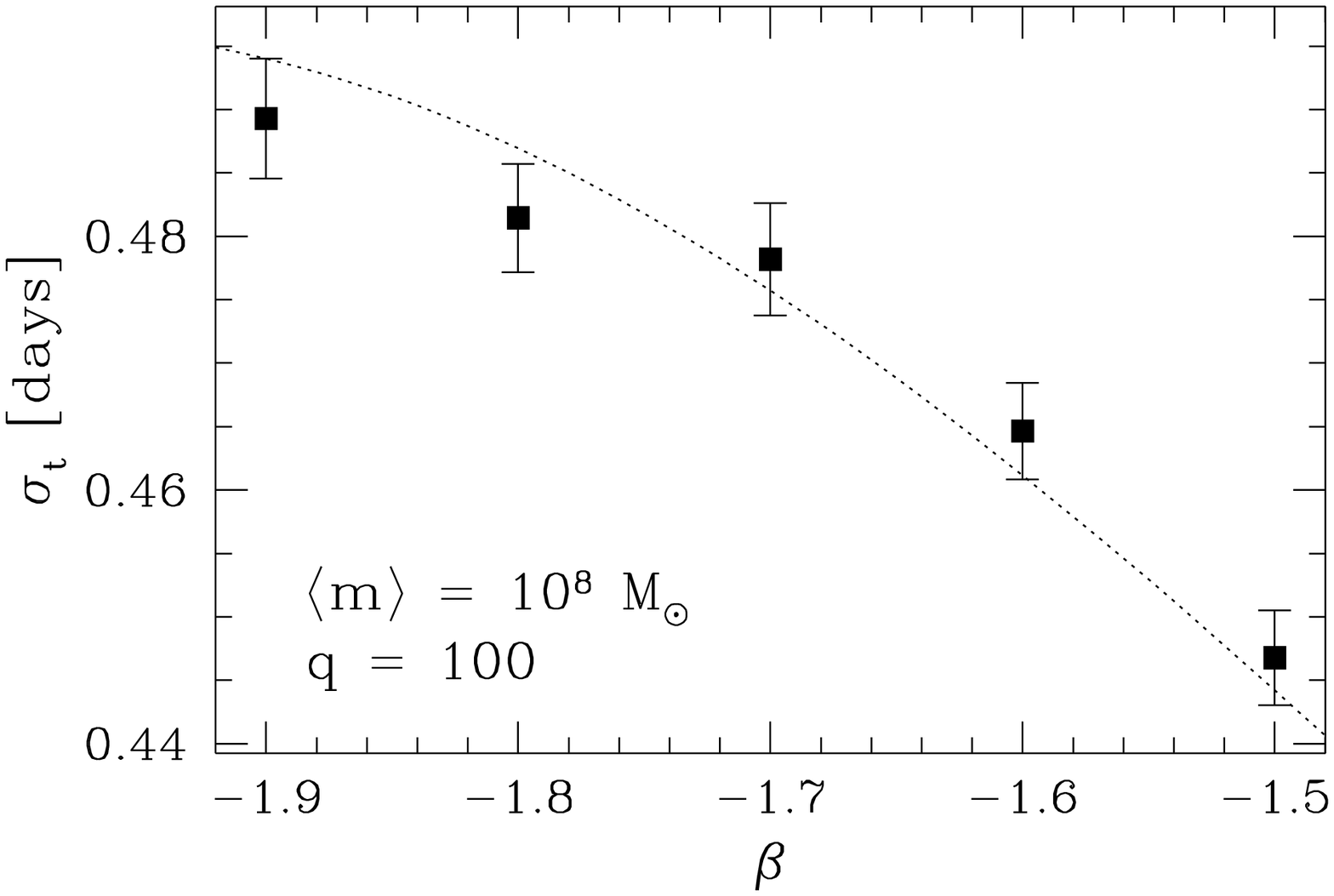}
}
\caption{
Dependence of the time delay scatter on the subhalo mass function.
The mass function is a power law, $dN/dm \propto m^\beta$, over the
range $m_1 \le m \le m_2$.  The two panels show $\sigma_t$ as a
function of the dynamic range $q = m_2/m_1$ and the power law slope
$\beta$.  The points show results from $10^4$ Monte Carlo simulations,
with bootstrap errorbars.  The dotted lines show the scalings
predicted by \refeq{sigf1} in the Appendix (after we fit for the
proportionality constant).  The substructure mass fraction is
$f_s = 0.01$.
}\label{fig:mfunc}
\end{figure*}

We can quantify this last point by plotting the time delay scatter
$\sigma_t$ as a function of the subhalo mass, for different values
of the substructure mass fraction, as shown in \reffig{mscaling}.
In the Appendix we use the Central Limit Theorem to make an initial
prediction of the scaling of the time delay scatter (see
\refeq{sigf2}):
\begin{equation}
\sigma_t \propto (f_s\,m)^{1/2} .  \label{eq:sigmat}
\end{equation}
To test this prediction, we use the results from simulations with
$f_s = 0.01$ to fit for the proportionality constant, and then plot
the analytic curves alongside the simulation results in
\reffig{mscaling}.  The agreement is striking given the simplicity
of the analytic prediction.  Strictly speaking, the analytic scaling
appears to overestimate the time delay scatter when the subhalo mass
is large.  We can identify at least two possible explanations.
First, when $f_s$ is fixed increasing the subhalo mass decreases the
number of subhalos, which may make the Central Limit Theorem less
applicable and invalidate an approximation in \refeq{sigphi}.
Second, as the subhalo mass increases the perturbations to the image
positions also increase, which may invalidate one of our starting
assumptions in the Appendix.  We are working on a more sophisticated
theory of time delay perturbations to address both issues.  For now,
though, we consider the success of the initial analytic prediction
to be very encouraging.

The scaling of the time delay scatter with subhalo mass has an
important corollary: there is no measurable effect on time delays
from low-mass objects, such as stars.  This means we can study
millilensing at any wavelength without ``contamination'' from
microlensing. 

\subsection{Subhalo mass function}
\label{sec:mfunc}

So far we have assumed that all subhalos have the same mass; now we
consider a power law mass function.  \reffig{mfunc} shows the time
delay scatter as a function of the dynamic range $q = m_2/m_1$ and
power law slope $\beta$ of the mass function.  The dynamic range has
a significant effect: the broader the mass function, the more scatter
there is in lens time delays.  The power law slope affects the time
delay scatter as well, but much more modestly.  For comparison we
also plot the analytic scaling from \refeq{sigf1} in the Appendix,
\begin{equation}
  \sigma_t \propto \left(f_s\,\frac{\avg{m^2}}{\avg{m}} 
    \right)^{1/2} ,
\end{equation}
where the ratio $\avg{m^2}/\avg{m}$ can be found from \refeq{msq}.
The analytic formula does not match the simulation results perfectly,
but it seems remarkably good for something so simple.  We take this
as an indication that the theory of time delay millilensing will be
analytically tractable.  Having a rigorous theory of millilensing
is possible only with time delays; analytic results for flux ratio
millilensing are elusive because flux ratios are highly nonlinear
and the Central Limit Theorem does not apply (see the Appendix).

\subsection{Internal structure of subhalos}
\label{sec:internal}

\begin{figure}[t]
\epsscale{1.0}
\plotone{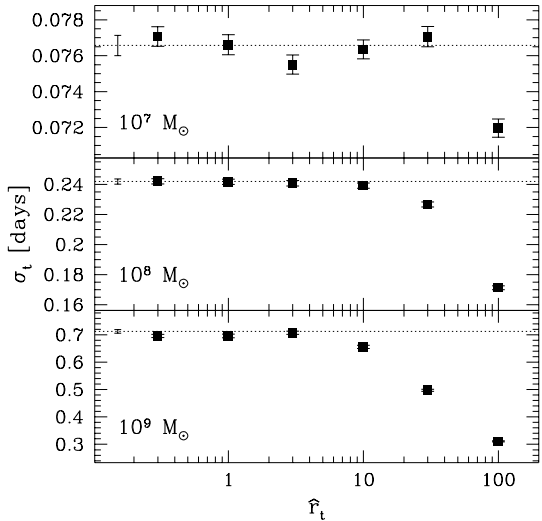}
\caption{
Time delay scatter for subhalos modeled as truncated isothermal
spheres, as a function of the dimensionless truncation radius
${\hat r}_t = r_t/\Rein$.  The dotted lines (with errorbars) show
the results for point mass subhalos for comparison.  The three panels
show different subhalo masses; the substructure mass fraction is
$f_s = 0.01$ in each case.  We show results for the M1/S2 image
pair, but the results for other pairs are similar.
}\label{fig:tsis}
\end{figure}

So far we have treated the mass clumps as point masses, partly for
simplicity and partly because we know the gravity outside any
spherical mass clump is the same as that of a point mass.  However,
dark matter subhalos in general have some spatial extent, and if
they overlap lensed images this may be important for millilensing.
In this pilot study we mainly want to determine when the point mass
approximation is reasonable and when we need to worry about subhalo
size and structure.  We therefore consider a simple model of
isothermal spheres that are truncated (in three dimensions) at some
radius $r_t$.
\reffig{tsis} shows how the time delay scatter changes when we vary
the truncation radius while keeping the subhalo mass fixed.  (The
key scale is the dimensionless truncation radius,
${\hat r}_t = r_t/\Rein$, where $\Rein$ is the Einstein radius of
a point mass with the same mass, as defined by \refeq{mhat}.)

Depending on the mass, subhalos need to extend beyond 10 or more
Einstein radii before the internal structure becomes important.  
The obvious next step is to model tidal truncation and estimate
the sizes of realistic subhalos.  There are some important effects
to consider: tidal forces vary with position in the parent halo;
and tidal truncation also corresponds to mass loss (so, strictly
speaking, it is not right to vary the truncation radius while
keeping the subhalo mass fixed, although that is a useful
pedagogical exercise).  These effects are incorporated into
semi-analytic substructure models, so in follow-up work we will
study time delay millilensing using realistic subhalo populations
drawn from those models.

\subsection{Arrival-time ordering of lensed images}
\label{sec:ordering}

\citet{SW-tdel} suggested that the arrival-time ordering of
lensed images is a robust prediction of smooth lens models (at
least, those with relatively simple angular structure).  With
the help of a few simple rules, they noted, it is usually easy
to identify the two minima and two saddlepoints in a quad lens,
and then to deduce the arrival-time ordering.  If the situation
is not immediately obvious, a simple lens model makes it clear.
We have therefore chosen to use the image classification and
ordering as the basis of our labeling scheme: in arrival-time
order we have the leading minimum (M1), the trailing minimum
(M2), then the leading saddle (S1), and finally the trailing
saddle (S2).

\begin{figure}[t]
\epsscale{1.0}
\plotone{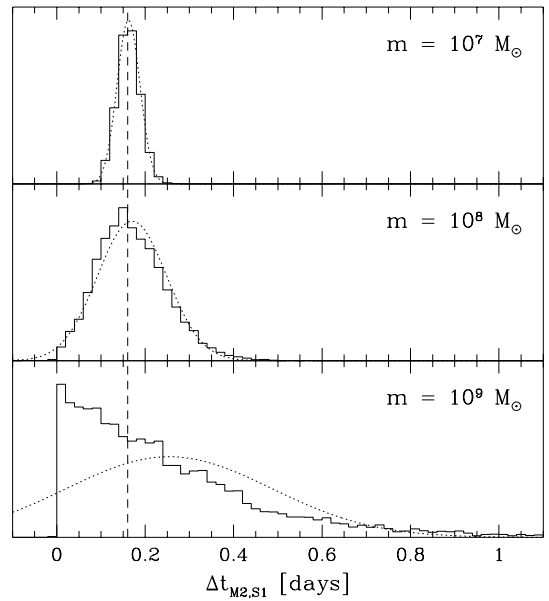}
\caption{
Histograms of the time delay of the close image pair M2/S1 for our
sample fold lens (compare Figs.~\ref{fig:hist} and \ref{fig:hist-m}).
The substructure mass fraction is $f_s = 0.01$, and the subhalo mass
is indicated in each panel.  The vertical lines indicate the time
delay for the reference smooth model.  Superposed on each histogram
is a Gaussian (dotted line) with the same mean and variance as the
simulated data.
}\label{fig:histBC}
\end{figure}

In the fold image configuration we have considered, substructure
does not affect the arrival-time ordering.  The reason is that the
time delay perturbations are small compared with most of the time
delays themselves.  It might seem that the close image pair M2/S1
could provide an exception, because the smooth model time delay
(0.16 day) is shorter than some of the substructure perturbations
we have seen.  However, by definition a saddlepoint in the time
delay surface must be higher than a nearby local minimum, so
image S1 must trail M2 and the overall ordering must be preserved.
One consequence is that the histogram for the M2/S1 time delay
cannot remain Gaussian as the substructure perturbations grow,
as shown in \reffig{histBC}.  This implies that the substructure
influences both images at some level, although the $\Delta t_{M2,S1}$
histograms are broad enough that it seems fair to say that
substructure affects even nearby images quite differently.  The
only way this general analysis can be violated is if a massive
and concentrated clump lies close enough to images M2 and S1 to
change the global topology of the time delay surface; but such a
situation would be immediately apparent from the image configuration
\citep[see][]{SW-tdel}.

\begin{figure*}[t]
\centerline{
  \includegraphics[width=2.5in]{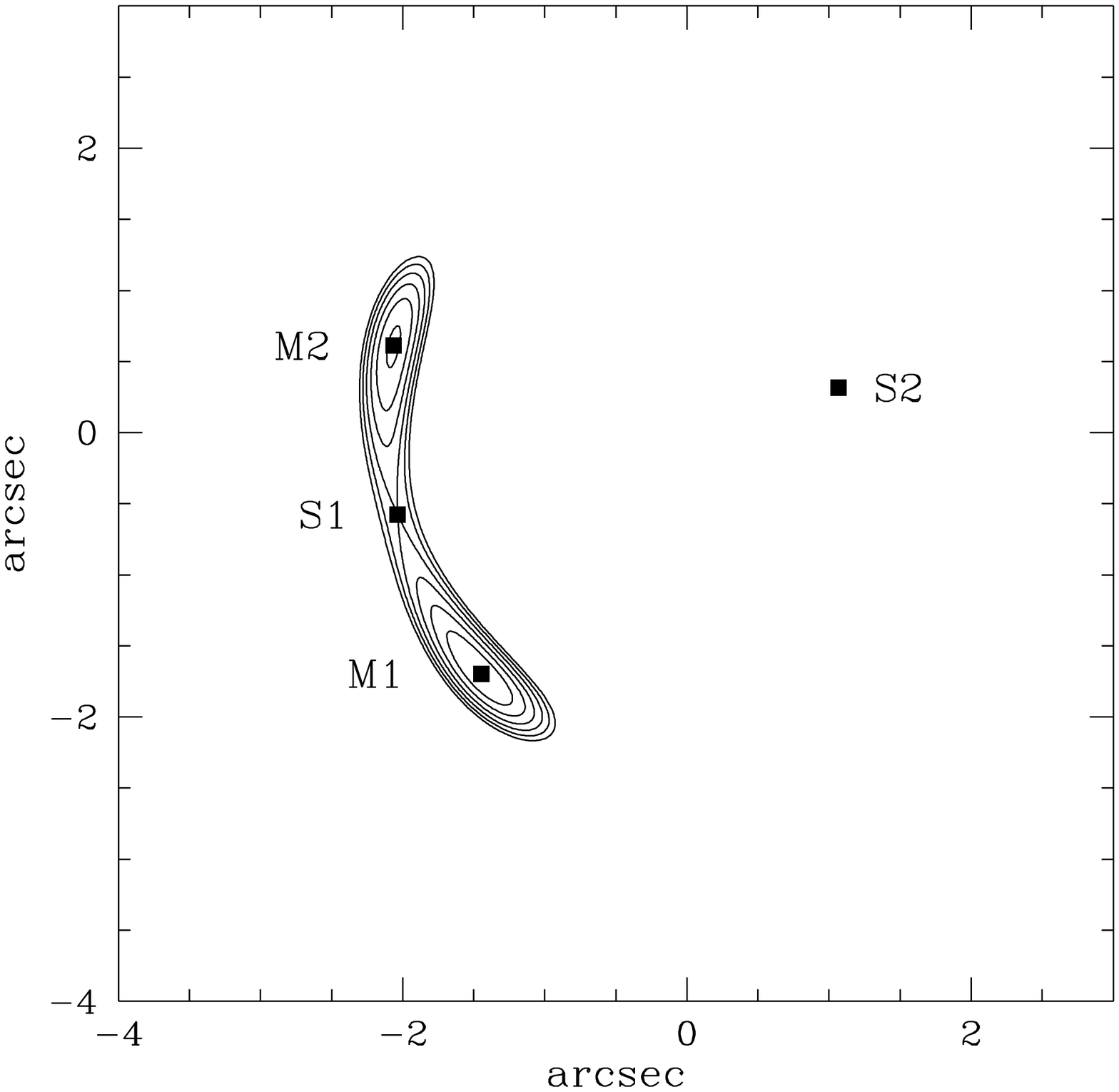}
  \includegraphics[width=2.5in]{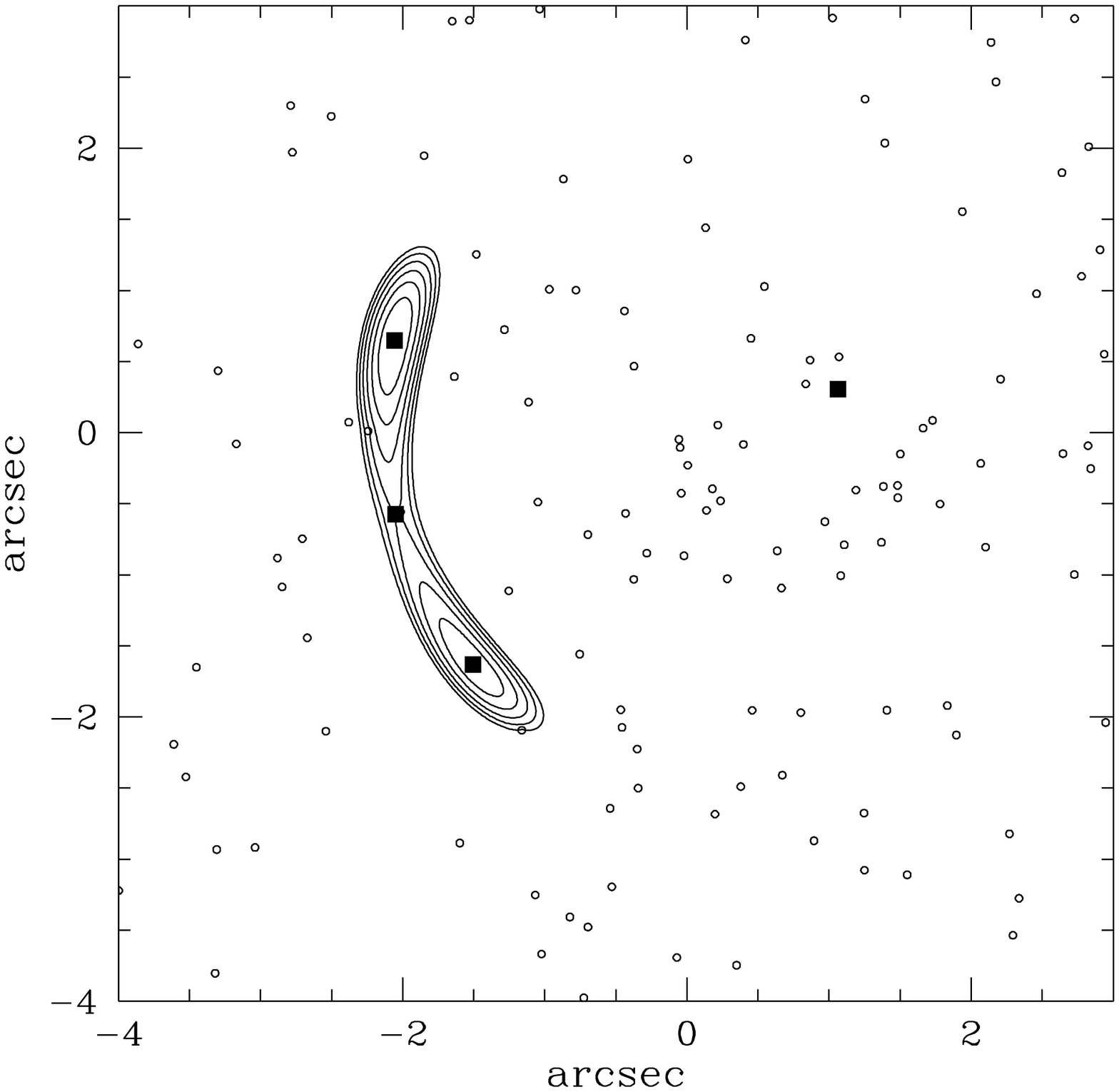}
}
\caption{
(a)
Image configuration for our sample cusp lens, assuming a smooth mass
distribution (compare with \reftab{cuspimg}).  In terms of the traditional
image labeling for RX J1131$-$1231, we have M1=C, M2=B, S1=A, and S2=D.
Positions are measured with respect to the center of the lens galaxy
at $(0,0)$.  Selected contours of constant arrival time are also
shown.  Counting contours shows that the minimum containing image
M2 is shallower than the minimum containing image M1, hence M2
trails M1.
(b)
Example of the same lens with substructure (with $f_s = 0.01$ and
$m = 10^{8}\,M_\odot$).  Now it is apparent that the minimum 
containing the image labeled M2 is actually deeper than the minimum
containing M1.  In other words, substructure has reversed the
arrival-time ordering of these two images such that M2 actually
precedes M1.  The ordering of S1 and S2 remains unchanged.
(Note that the contour levels are different in the two panels;
they are chosen to best reveal the minima and saddlepoint in each
case.)
}\label{fig:cuspimg}
\end{figure*}

\begin{figure}[t]
\epsscale{1.0}
\plotone{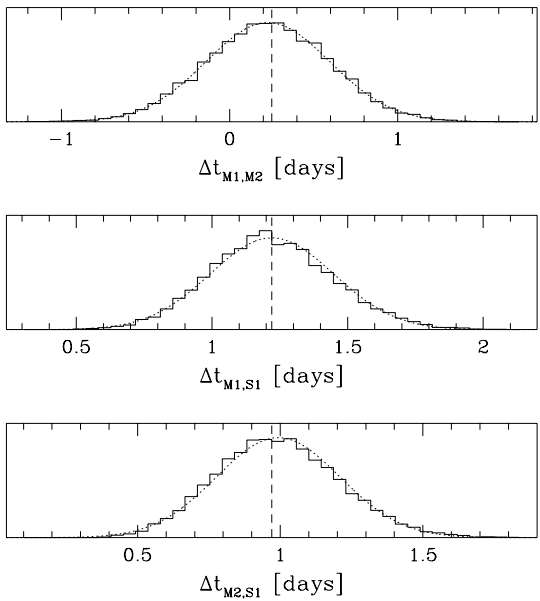}
\caption{
Histograms of the time delays between the close images in our
example cusp lens.  The images labeled M1 and M2 lie at minima of
the time delay surface, while S1 lies at a saddlepoint.  In smooth
models the image ordering is always M1/M2/S1/S2.  However, in Monte
Carlo simulations with a substructure mass fraction $f_s = 0.01$
and subhalo mass $m = 10^8\,M_\odot$, about 27\% of the realizations
have $\Delta t_{M1,M2} < 0$ which means image M2 actually leads
image M1.  In such cases the overall ordering in M2/M1/S1/S2.  As
usual, vertical lines indicate the time delays for the reference
smooth model, and a comparison Gaussian is superposed on each
histogram.
}\label{fig:hist1131}
\end{figure}

The issue of arrival-time ordering is strikingly different for a
cusp image configuration like that in RX J1131$-$1231.
\reffig{cuspimg}(a) shows the images along with selected arrival-time
contours for our smooth mass model.  In this example, and indeed
in all smooth models of RX J1131$-$1231 examined by \citet{1131tdel},
the image labeled M1 is the leading image.  However, as shown in
\reffig{cuspimg}(b), substructure can reverse the arrival-time order
of the images M1 and M2, such that the overall ordering is
M2/M1/S1/S2.  In terms of the topology of the time delay surface,
we can still say the saddlepoint must be higher than the two minima
on either side, so S1 must trail both M1 and M2.  However, there
are no such restrictions on the heights (or depths, equivalently)
of the two minima with respect to each other.  It is fairly easy
for substructure to modify one or both of the minima enough to
change their relative heights, and hence the arrival-time ordering
of images M1 and M2.  In fact, some 27\% of realizations with modest
substructure (substructure mass fraction $f_s = 0.01$ and subhalo
mass $m = 10^{8}\,M_\odot$) predict an arrival-time reversal (see
\reffig{hist1131}).

We believe image reversal by substructure may already have been
observed in RX J1131$-$1231, based on the time delays
reported by \citet{1131tdel}.  The time ordering observed among
the three bright images is M2/M1/S1\footnote{We have switched to
our labeling scheme using the identifications M1=C, M2=B, S1=A,
and S2=D.} with
$\Delta t_{M2,M1} = 2.20_{-1.64}^{+1.55}$ days,
$\Delta t_{M2,S1} = 11.98_{-1.27}^{+1.52}$ days, and
$\Delta t_{M1,S1} = 9.61_{-1.57}^{+1.97}$ days.
These time delays present two puzzles.  First, image M2 is observed
to lead image M1, whereas smooth models predict the opposite.  Second,
the observed M2/S1 and M1/S1 time delays are an order of magnitude
longer than predicted by smooth models.  \citet{1131tdel} focused
their attention on the second problem, and showed that placing a
single, massive ($\gtrsim 5 \times 10^{10}\,M_\odot$) subhalo near
image S1 could explain the long M2/S1 time delay.  Their models
still predicted M1 to be the leading image, though.

As we have demonstrated, a {\em population} of subhalos might be
able to reverse the ordering to make M2 the leading image.  It will
still take a lot of modeling to draw quantitative conclusions about
the substructure required to explain the image ordering and the long
time delays; we will present the modeling details elsewhere (C.~R.~Keeton
\& L.~A.~Moustakas, in preparation).  For now, we mainly want to introduce the
idea that the temporal ordering of images in a cusp configuration
can be changed by substructure---and that this effect may already
have been observed.

\subsection{Macromodel uncertainties}
\label{sec:macro}

Having shown that substructure affects lens time delays, we still
need to consider how well the perturbations could be detected in
observed lenses.  The question is whether uncertainties in the
``macromodel'' used to characterize the smooth component of the
lens mass distribution could mask the effects of substructure.
One key concern is the radial profile degeneracy: especially in
quad lenses, varying the radial density profile of the lens galaxy
can rescale the time delays while leaving other observables unchanged
\citep[e.g.,][]{FGS,kk1115,saha-degen,cskH0}.  At first glance,
this would seem to make changes induced by substructure degenerate
with changes in the radial profile.

\begin{figure*}[t]
\centerline{
  \includegraphics[width=1.9in]{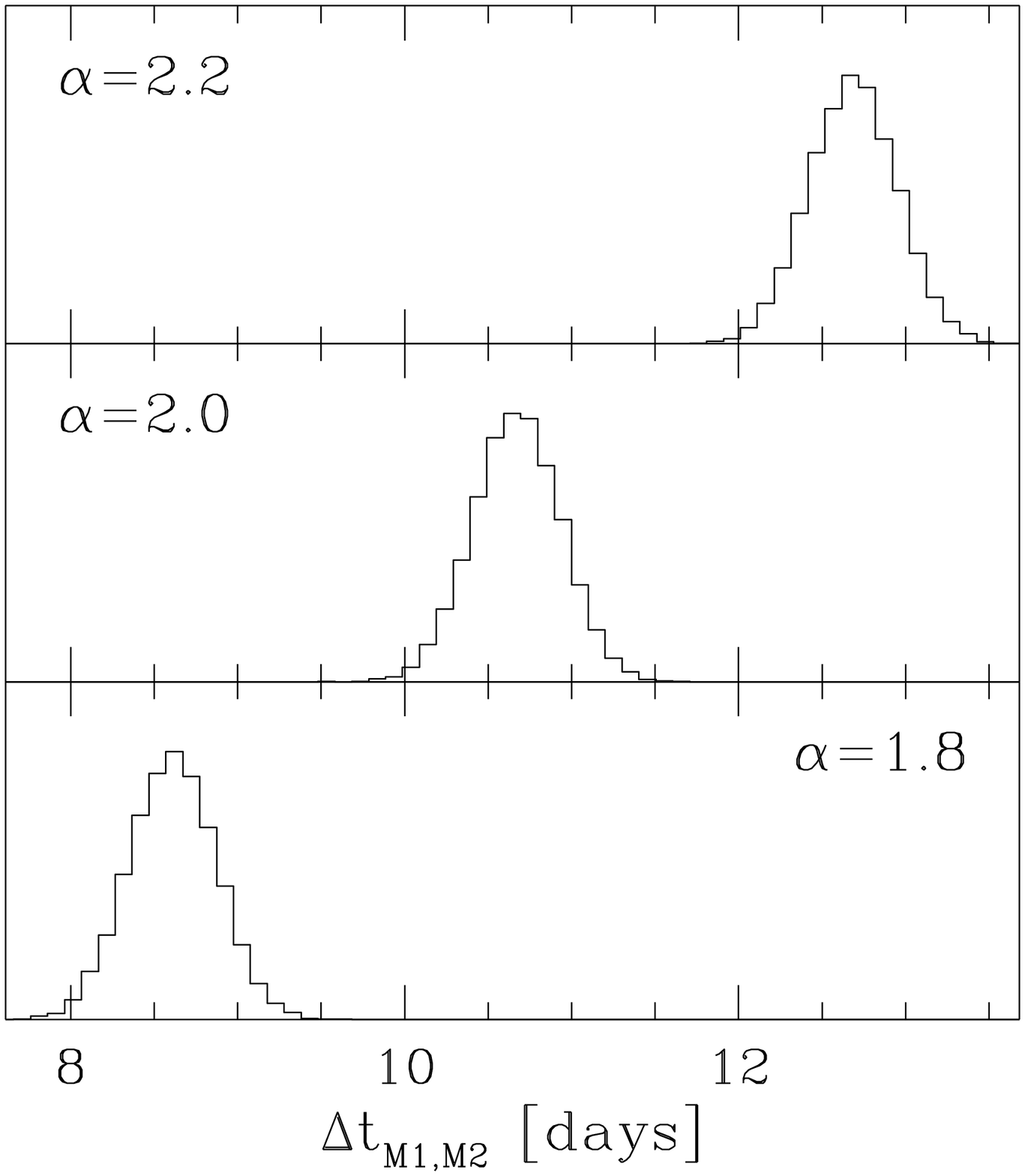}
  \includegraphics[width=1.9in]{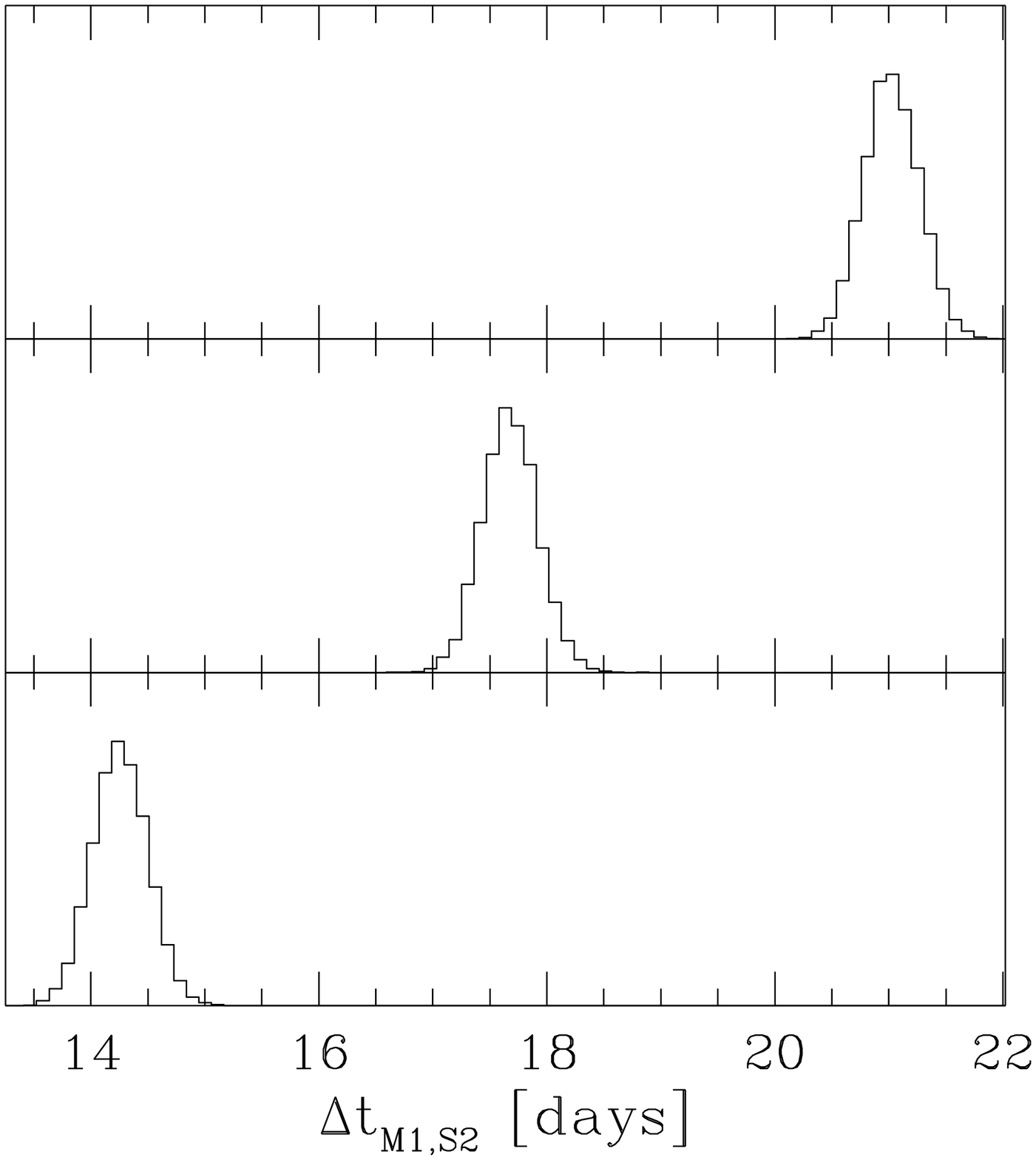}
  \includegraphics[width=1.9in]{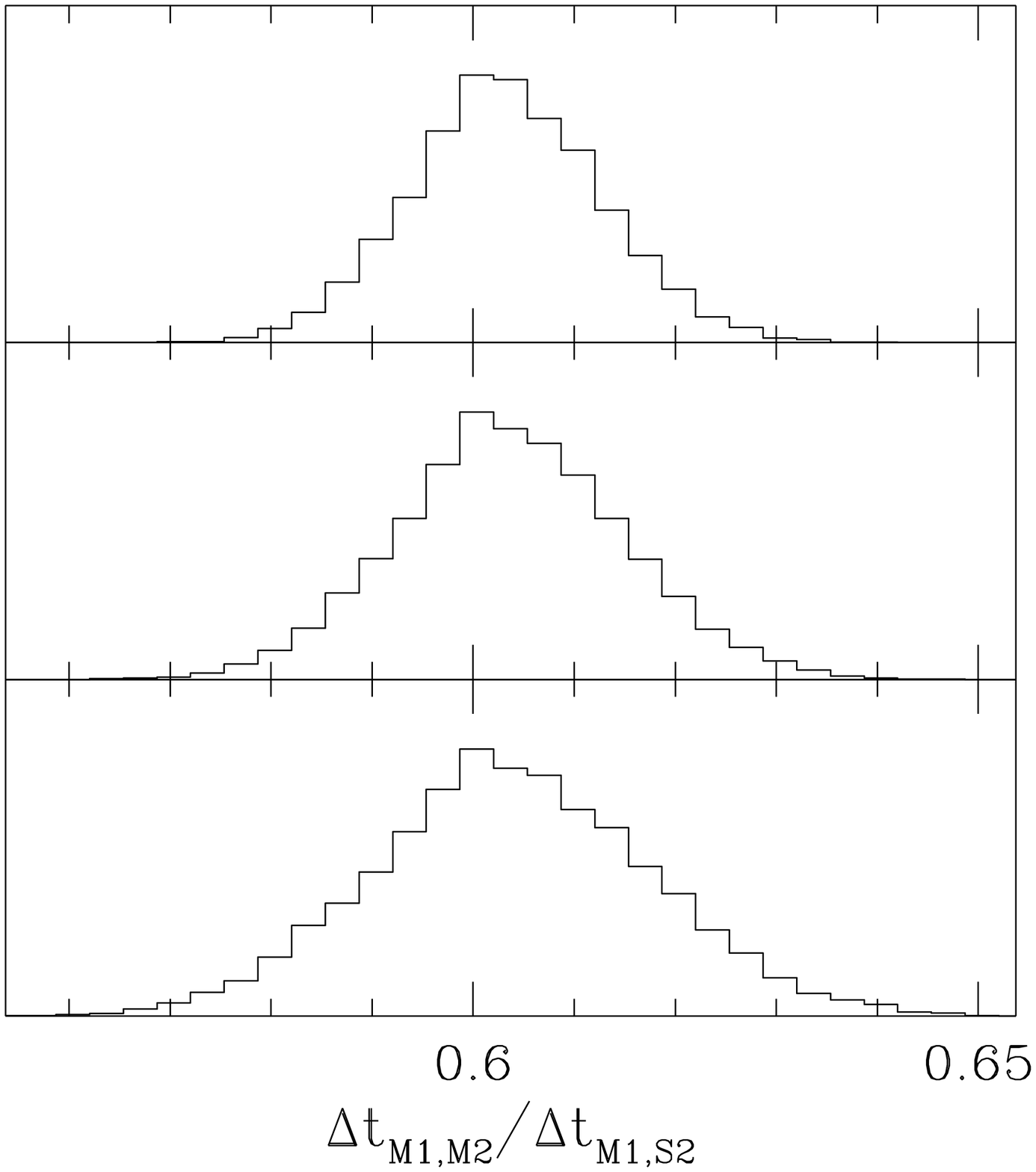}
}
\caption{
Examining the radial profile degeneracy in our mock fold lens.
The smooth mass component has a power-law density profile
$\rho \propto r^{-\alpha}$, or equivalently a surface density
profile $\kappa \propto R^{1-\alpha}$, while the substructure
is still given by our pseudo-Jaffe model.  Varying the profile
rescales the time delays (left and middle columns).  However,
the time delay {\em ratio} (right column) is largely unaffected:
the mean ratio stays the same, while there is a small change in
the scatter.  (The time delays between images M1, M2, and S2 are
used for illustration; time delays involving image S1 could be
used as well.)  Time delay ratios therefore allow us to probe
dark matter substructure without worrying about the radial
profile degeneracy.
}\label{fig:alpha}
\end{figure*}

\begin{figure*}
\epsscale{0.7}
\plotone{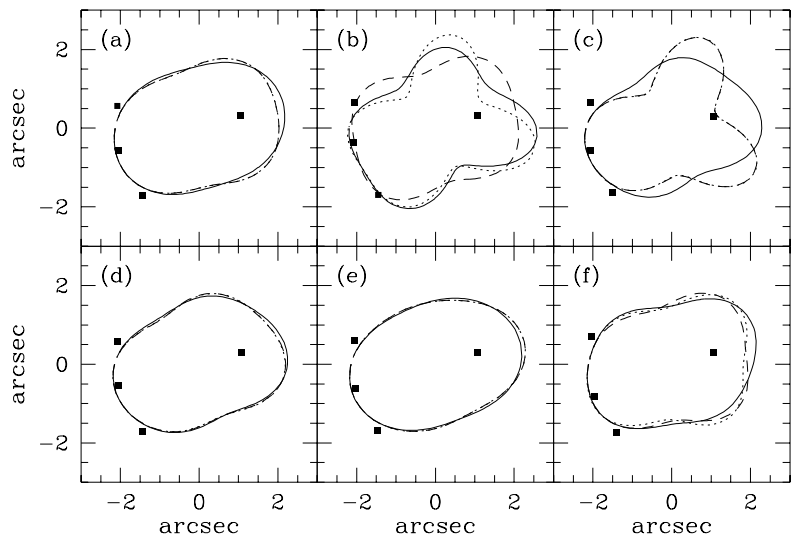}
\caption{
Lensing critical curves for multipole models fit to six random
mock lenses.  We fit the image positions and time delays using
multipole terms that are unconstrained (dotted curves) or have
Gaussian priors (solid curves; see text).  We also consider
adding the image flux ratios as constraints (dashed curves).
Among these lenses, cases b and c are the most glaring examples
of failed multipole models; cases d and f are less dramatic but
failures nonetheless.
}\label{fig:mpole}
\end{figure*}

However, the radial profile rescaling affects all images in the
same way, while substructure affects each image differently.  So
with more than one time delay (as in a quad lens), ratios of the
delays cancel the rescaling and isolate substructure effects, as
shown in \reffig{alpha}.  The radial profile degeneracy will
therefore not be a fundamental limitation in attempts to probe
substructure with time delays.  This very promising inference is
currently empirical, but we expect it can be made rigorous with
a full theory of time delay millilensing.  Since radial profile
effects do not cancel perfectly (there are small differences in
the widths of the time delay ratio histograms in \reffig{alpha}),
the theory will also be useful in understanding and correcting
for these residual effects.

Another source of uncertainty is the angular structure of the
macromodel.  While it is common to fit observed lenses with an
ellipsoidal mass distribution and external shear (and we have used
such a model to construct our mock lenses), real galaxies may be
more complex.  In particular,
observed \citep[e.g.,][]{bender,saglia,rest}
and simulated \citep[e.g.,][]{heyl,NB03,JNB05,naab06}
elliptical galaxies often contain mild departures from elliptical
symmetry that can be described with $m=3$ and $m=4$ multipole
terms in the surface mass density.  \citet{EW} suggested that
such multipole terms might provide an alternative to clumpy
substructure when fitting observed lenses.  Although multipoles
turn out not to provide a satisfactory explanation of observed
flux ratio anomalies \citep{KD,congdon-mpole,yoo1,yoo2}, it is
important to consider how they might influence the analysis of
time delays.

We can do this by taking our mock lenses generated with
substructure and trying to fit them with multipole models.  For
this exercise we adopt $0\farcs003$ errorbars for the image
positions, which is typical of observations with the {\em Hubble
Space Telescope (HST)} and radio interferometry, and 5\% time delay
uncertainties, which is similar to the precision achieved for
quad lenses today \citep[see the compilation by][]{oguriH0}.
In general, we find that multipole models can formally provide
a good fit to the image positions and time delays, but many of
the models have unreasonably large multipole terms that lead to
unrealistic galaxy shapes, as shown in \reffig{mpole}.  To avoid
the unreasonable models we can impose some priors on the multipole
terms.  Since the multipole moments in observed and simulated
elliptical galaxies are typically at the percent level, a simple
first step is to adopt Gaussian priors on the multipole amplitudes
$a_3$ and $a_4$ with a mean of zero and standard deviation of
0.02.  (In the present analysis we do not impose any constraints
on the orientations of the multipole terms.)  With these mild
amplitude priors, multipole models fail (i.e., they are
inconsistent with the data at more than 95\% confidence) for 27
of the 100 mock lenses we examine.  We are in the process
(J.\ van Saders et al., in preparation) of deriving more sophsticated
priors on multipole terms from a large sample of galaxies in the
Galaxy Evolution from Morphologies and SEDs survey
\citep[GEMS;][]{GEMS}, and we expect that those will be even more
valuable in constraining macromodels and thus aiding the
identification of time delay anomalies.

It is perhaps artificial to insist that we must identify anomalous
lenses on the basis of time delays alone, when flux ratios are
also available to provide additional constraints.  If we fit the
flux ratios along with the image positions and time delays, we
find that multipole models fail (at 95\% confidence) for 52 of
100 mock lenses even if we use generous 10\% flux errorbars and
do not impose any priors on the multipole terms.  Interestingly,
only 28 of those failures would be identified on the basis of
flux ratios alone; in 24 of the mock lenses, it is only when we
fit the time delays and flux ratios simultaneously that the
anomalies become clear.  We infer that time delays and flux
ratios contain different and complementary information about the
small-scale structure of the lens potential.

While we plan to address the issues of multipole priors and
complementarity of time delays and flux ratios in detail in
future work, we offer the preliminary conclusion that macromodel
uncertainties can be controlled well enough that we can expect to
use time delays to learn about substructure.

\section{Discussion and Conclusions}
\label{sec:discuss}

We have shown that substructure in lens galaxies modifies the time
delays in multiply-imaged gravitational lenses.  The amplitude of
time delay perturbations depends on the abundance of substructure,
the mass function of subhalos, and to some extent the internal
structure of subhalos as well.  This phenomenon, which we call
time delay millilensing, has some very attractive properties.
First, the effects of substructure on time delays are fairly easy
to understand, both conceptually and quantitatively.  Second,
time delay perturbations are unaffected by stellar microlensing
or extinction in the lens galaxy.  Third, time delay ratios are
immune to the radial profile degeneracy in lens modeling.

Furthermore, time delay millilensing is sensitive to the mass
function of subhalos in a different way than flux ratio and
astrometric millilensing.  For point mass perturbers, the
lensing optical depth depends on
$\int \Rein^2\,(dN/dm)\,dm \propto \int m\,(dN/dm)\,dm$.  The
potential of a clump scales with its mass, so time delay
perturbations depend in full on $\int m^2\,(dN/dm)\,dm$.  (See
the Appendix for more details.) For comparison, position
perturbations have a scale of $\Rein \propto m^{1/2}$ while
magnification perturbations are dimensionless, which means
that astrometric millilensing depends on
$\int m^{3/2}\,(dN/dm)\,dm$ while flux ratio millilensing
depends on $\int m\,(dN/dm)\,dm$
\citep[compare][]{DK,dobler1,chenastro,shin}.

We therefore suggest that lens time delays should join image
positions and flux ratios as tools for studying dark matter
substructure in distant galaxies.  Since the three observables
measure different moments of the subhalo mass function, the
combination of all three will ultimately provide a powerful
way to test CDM predictions and even to explore the nature of
dark matter itself, given that different dark matter candidates
may produce different mass functions at subgalactic scales, as
a function of lens galaxy mass and redshift.

Before we try to use time delays to measure substructure, it is
vital that we understand whether systematic uncertainties in lens
models could corrupt substructure constraints.  We have explicitly
shown that the radial profile degeneracy is not a significant
problem, as time delay ratios essentially factor out the profile
dependence.  We have drawn the preliminary conclusion that
uncertainties in the angular structure of the lens potential can
be dealt with, using additional flux ratios data and/or independent
constraints on the shapes of galaxy mass distributions.  Other
potential concerns may include the environment of the lens galaxy
\citep{KZ04}, and (sub)structure along the line of sight
\citep{anal,chenLOS,metcalfLOS1,metcalfLOS2,lieu}.  For all of
these issues, the important point is again that non-local features
in the lens potential affect the images in some coordinated way,
whereas substructure affects the images differently.  

We also need to predict time delay perturbations for subhalo
populations that are more realistic than the simple ones we have
used (for pedagogical purposes) here.  This work is underway, using
existing semi-analytic substructure models.  By incorporating
important physical effects such as accretion of new subhalos and tidal
truncation and disruption of old subhalos, these models yield subhalo
populations whose spatial distributions and mass functions agree well
with results from $N$-body simulations
\citep[e.g.,][]{TB01,TB04,benson,ZB03,koush,oguri-sub,vdb,Z05}.
Our follow-up calculations will offer greater insight into the
subtleties of time delay perturbations in a broader range of
realistic circumstances, and better guidance as to how well we need
to measure time delays if we want to probe dark matter substructure.

We have noted that in extraordinary lenses with close triplets of
images in a cusp configuration, it may be possible to detect
the effects of substructure through unambiguous changes in the
arrival-time ordering of the images (compared to expectations for
standard macromodels).  We expect, though, that the real power of
time delay millilensing will lie in its application to large
ensembles of lenses with measured time delays.  At present, time
delay uncertainties are typically at the level of 1 day (see
\citealt{oguriH0} for a compilation).  Improvements of a factor
of a few are desired: our preliminary estimates indicate that
improving the uncertainties to $\sim$0.5\% would yield constraints
on substructure at the level of 0.3--0.5 dex in a given 4-image
lens.  \citet{colley} have shown that coordinated global optical
monitoring can yield time delays with uncertainties of less than
0.1 day.  Monitoring at X-ray wavelengths, where variability can
be rapid and dramatic, can yield even more precise time delays:
\citet{chartas} measure a time delay of $3.58\pm0.14$ hours
between the two close images in PG 1115+080 with {\em Chandra}
and {\em XMM-Newton} observations.  With some effort, such techniques can
probably be extended to a modest sample of time delays in the
near future.  Over the longer term, the advent of large time-domain
surveys is expected to yield thousands of lensed quasars with
time delays \citep[e.g.,][]{LSSTlens,kuhlen,SNAPlens,csk-find}.
Even better would be a dedicated space platform capable of
precise monitoring of lenses over extended periods of time
\citep[see][]{omega}.  Looking to the future, measuring precise
image positions, flux ratios, and time delays in a sample of
$\sim$100 quad lenses could yield substructure measurements at
the 20\%--30\% level or better.  Since the measurements will span
the redshift range $0.2 \lesssim z \lesssim 1$, this will provide
a unique opportunity to study not just the amount of dark matter
substructure but its evolution as well.

\acknowledgements 
We thank Michael Kuhlen, George Meylan, Arlie Petters, and
Eduardo Rozo for helpful conversations.  We thank Masamune Oguri
and the referee Neal Dalal for raising the issue of modeling
uncertainties associated with multipole terms.
CRK thanks Jonathan Faiwiszewski for suggesting that the Central
Limit Theoreom might be helpful.
CRK is supported by grant HST-AR-10668 from the Space Telescope
Science Institute, which is operated by the Association of
Universities for Research in Astronomy, Inc., under NASA contract
NAS5-26555; and by NSF grant AST-0747311.  The work of LAM was
carried out at Jet Propulsion Laboratory, California Institute
of Technology, under a contract with NASA.

\appendix

\section{Statistics of Time Delay Perturbations}

In this Appendix we present an initial theory of the statistics
of time delay perturbations.  This theory rests on three
assumptions:
(1) subhalos are statistically independent (of each other; the
probability of having a subhalo may still depend on position);
(2) subhalos may be treated as point masses; and
(3) time delay perturbations are dominated by changes in the
lens potential (as opposed to changes in the image positions).
The first assumption means that all subhalo positions and
masses are drawn from the same probability distribution.  This
simplifies the theory, and seems reasonable for lensing since
we work in projection and subhalos that are near each other
on the sky are most likely well separated in space.  The second
and third assumptions will be relaxed in future work, but are
adequate for our present goal of understanding in a general way
how the time delay scatter depends on the amount of substructure
and the subhalo mass function.

The lens time delay has the form given by \refeq{tdel}.  By
assumption \#3, if we want to understand how substructure affects
$\Delta t_{12} = \tau(\xv_1) - \tau(\xv_2)$, it is sufficient
to study the potential difference
$\Delta\phi_{12} = \phi(\xv_1) - \phi(\xv_2)$.  Any change in
the time delay is simply given by the change in the potential
difference, multiplied by the time scale $t_0$.

The potential due to a collection of $N$ subhalos is the sum of
contributions from individual subhalos.  If the subhalos are
point masses, the potential difference has the form
\begin{equation}
  \Delta\phi_{12} = \sum_{i=1}^{N} \varphi(\rv_i,m_i)\,,
  \quad\mbox{where}\quad
  \varphi(\rv,m) = \frac{\mhat}{\pi} \ln \frac{|\xv_1-\rv|}{|\xv_2-\rv|}\ .
\end{equation}
If we can show that the mean and variance of $\varphi$ are
finite, then we can use the Central Limit Theorem to infer
that the sum $\Delta\phi_{12}$ will be approximately Gaussian,
and to determine the variance of $\Delta\phi_{12}$.

In order to compute the mean and variance of $\varphi$, we need
to average over a subhalo's position and mass.  Let $p_r(\rv)$
be the probability distribution for the subhalo's position,
while $p_m(m) = (1/N)\,dN/dm$ is the probability distribution
for its mass.  Then the average of any function $f$ has the form
\begin{equation}
  \avg{f} \equiv \int dm\ p_m(m) \int d\rv\ p_r(\rv)\ f\,.
\end{equation}
It is useful to note that smoothed substrucure density
distribution, $\ks(\rv)$, may be written in terms of the
position probability distribution:
\begin{equation} \label{eq:ks}
  \ks(\rv) \equiv \int dm\ \frac{dN}{dm}\ \pi\Rein^2\ p_r(\rv)
  = \int dm\ \frac{dN}{dm}\ \mhat\ p_r(\rv)
  = N \avg{\mhat} p_r(\rv)\,.
\end{equation}
Now in order to determine the mean and variance of $\varphi$,
we need to compute the moments
\begin{eqnarray}
  \avg{\varphi} &\equiv& \int dm\ p_m(m) \int d\rv\ p_r(\rv)\ 
    \frac{\mhat}{\pi}\ \ln \frac{|\xv_1-\rv|}{|\xv_2-\rv|}
\nonumber\\
  &=& \frac{1}{N \pi} \int d\rv\ \ks(\rv)\ 
    \ln \frac{|\xv_1-\rv|}{|\xv_2-\rv|}\ ,
\label{eq:phi1int} \\
  \avg{\varphi^2} &\equiv& \int dm\ p_m(m) \int d\rv\ p_r(\rv)\ 
    \frac{\mhat^2}{\pi^2} \left(\ln \frac{|\xv_1-\rv|}{|\xv_2-\rv|}\right)^2
\nonumber\\
  &=& \frac{1}{N \pi^2}\ \frac{\avg{\mhat^2}}{\avg{\mhat}}
    \int d\rv\ \ks(\rv)\ \left(\ln \frac{|\xv_1-\rv|}{|\xv_2-\rv|}\right)^2 .
\label{eq:phi2int}
\end{eqnarray}
In both cases we used \refeq{ks} to substitute for $p_r(\rv)$ in
terms of $\ks$.  We also identified the mass integrals as yielding
the averages $\avg{\mhat}$ and $\avg{\mhat^2}$.
The variance is then
\begin{equation} \label{eq:sigphi}
  \sigma_\varphi^2 \ =\ \avg{\varphi^2} - \avg{\varphi}^2
  \ \approx\ \avg{\varphi^2}
  \ \approx\ \frac{1}{N \pi^2}\ \frac{\avg{\mhat^2}}{\avg{\mhat}}
    \int d\rv\ \ks(\rv)\ \left(\ln \frac{|\xv_1-\rv|}{|\xv_2-\rv|}\right)^2 .
\end{equation}
At the second step, we note that $\avg{\varphi^2} \sim 1/N$
while $\avg{\varphi}^2 \sim 1/N^2$, so when there are many
subhalos we may neglect the second term.

The integrands in \refeqs{phi1int}{phi2int} diverge near $\xv_1$
and $\xv_2$, but only logarithmically, and such a divergence is
still integrable.  As a result, as long as $\ks(\rv)$ does not
diverge badly near the image positions, the integrals are well
behaved, so the mean and variance of $\varphi$ are finite.  This
means the Central Limit Theorem holds, and we can infer that the
potential difference $\Delta\phi_{12}$, and hence the time delay
$\Delta t_{12}$, will have a distribution that is approximately
Gaussian.  (This argument also applies to clumps that are not
point masses, because then any divergence in $\varphi$ will be
softer than logarithmic.)

Furthermore, thanks to the Central Limit Theorem we may compute
the variance in $\Delta\phi_{12}$ as the quadrature sum of the
variances of all the individual $\varphi$ terms---or, in this
case, a simple multiplication by $N$:
\begin{equation} \label{eq:siggen}
  \sigma_\phi^2 \ \approx\ N\,\sigma_\varphi^2
  \ \approx\ \frac{1}{\pi^2}\ \frac{\avg{\mhat^2}}{\avg{\mhat}}
    \int d\rv\ \ks(\rv)\ \left(\ln \frac{|\xv_1-\rv|}{|\xv_2-\rv|}\right)^2 .
\end{equation}
And of course $\sigma_t = t_0\,\sigma_\phi$.  Heuristically, we
may summarize this result as
\begin{equation} \label{eq:sigk}
  \sigma_t \propto \left(\ks\,\frac{\avg{m^2}}{\avg{m}} 
    \right)^{1/2} ,
\end{equation}
where $\ks$ is indicative of the amount of substructure.  While
this expression oversimplifies the dependence on the spatial
distribution of subhalos, it is conceptually instructive.  In
our simulations, we write $\ks(\rv) = f_s\,\kappa_\tot(\rv)$,
and we keep the total mass distribution $\kappa_\tot$ fixed while
we change the substructure mass fraction $f_s$.  Therefore it is
formally correct to write
\begin{equation} \label{eq:sigf1}
  \sigma_t \propto \left(f_s\,\frac{\avg{m^2}}{\avg{m}} 
    \right)^{1/2} .
\end{equation}
This is our general conclusion for how the time delay scatter
depends on the amount of substructure and the subhalo mass
function.  Note that if all subhalos have the same mass $m$,
this simplifies to
\begin{equation} \label{eq:sigf2}
  \sigma_t \propto \left(f_s\,m\right)^{1/2} .
\end{equation}

It is instructive to consider whether the Central Limit Theorem
can be applied to flux ratios as well as time delays.  For point
mass clumps, the magnification of an image at position $\xv$
has the form
\begin{equation}
  \mu = \left[ (1-\phi_{xx})(1-\phi_{yy}) - \phi_{xy}^2 \right]^{-1} ,
\end{equation}
where
\begin{equation}
  \phi_{xx} = - \sum_{i=1}^{N} \frac{\mhat}{\pi}\ 
    \frac{(x-x_i)^2-(y-y_i)^2}{|\xv-\xv_i|^4}\ ,
\end{equation}
and there are similar expressions for $\phi_{yy}$ and $\phi_{xy}$.
Now when we try to compute the mean and variance of each term by
integrating over $\xv_i$, the integrand diverges non-integrably
near $\xv$, and so the mean and variance diverge
\citep[also see][]{WMS95}.  The Central Limit Theorem therefore
does not apply.  The magnification probability distribution
$p(\mu)$ may still be perfectly well defined, but it need not be
Gaussian.  Indeed, $p(\mu)$ can be computed analytically for a
uniform distribution of point masses, and it is distinctly
non-Gaussian \citep[see \S 11.2 of][]{SEF}.  There may be
circumstances (say, certain types of spatially extended clumps)
for which the Central Limit Theorem does apply to magnifications,
but the theorem is not universally applicable as it is for time
delays.

Returning to time delays, there is one last bit of theory that
is useful.  We intuitively expect that clumps far from the images
have little effect on the time delay, but we can quantify this
statement.  Consider for simplicity images at $\xv_1 = (d,0)$
and $\xv_2 = (-d,0)$, and let us examine only subhalos at radii
$r > R_0 \gg d$.  We return to \refeq{siggen}, plug in for
$\xv_1$ and $\xv_2$, and then take a Taylor series expansion to
lowest order in $d/r$:
\begin{equation} \label{eq:Rthresh}
  \epsilon_\phi^2 \ \approx\ \frac{1}{\pi^2}\ 
    \frac{\avg{\mhat^2}}{\avg{\mhat}} \int_{0}^{2\pi} d\theta
    \int_{R_0}^{\infty} dr\ r\ \ks(r)\ 4 d^2\ \frac{\cos^2\theta}{r^2}
  \ \approx\ \frac{4d^2}{\pi}\ \frac{\avg{\mhat^2}}{\avg{\mhat}} 
    \int_{R_0}^{\infty} \frac{\ks(r)}{r}\ dr\,.
\end{equation}
Notice that we have changed notation and written $\epsilon_\phi$ to
highlight that this is an error term---specifically, the rms error
we make by neglecting subhalos beyond $R_0$.  We have computed the
error in the potential, but we can relate it to the rms error in
the time delay via $\epsilon_t = t_0\,\epsilon_\phi$.  (We have
verified \refeq{Rthresh} using direct simulations of subhalo
populations extending to large radii.)  We have assumed for
simplicity that the substructure mass distribution is circularly
symmetric, $\ks(\rv) = \ks(r)$, but the generalization to elliptical
symmetry is straightforward.  For any reasonable substructure
distribution, $\ks(r)$ decreases as $r \to \infty$, so the integral
converges.  Not only that, but $\epsilon_t$ is a monotonically
decreasing function of $R_0$.  In other words, the farther the
subhalos are from the images, the less effect they have on the
time delays.  If we can tolerate some small error in time delays,
we may neglect all subhalos beyond some threshold radius $R_0$.
Once we specify the time delay error tolerance, we can solve
\refeq{Rthresh} to compute the threshold radius $R_0$.


\begin{thebibliography}{}

\bibitem[Amara et al.(2006)]{amara} 
Amara, A., Metcalf, R.~B., Cox, T.~J., \& Ostriker, J.~P.,  
2006, 
\mnras, 367, 1367

\bibitem[Barnes \& Hut(1986)]{tree} 
Barnes, J., \& Hut, P.,  
1986, 
\nat, 324, 446

\bibitem[Bender et al.(1989)]{bender}
Bender, R., Surma, P., D\"obereiner, S., M\"ollenhoff, C., Madejski, R.
1989, \aap, 217, 35

\bibitem[Benson et al.(2002)]{benson} 
Benson, A.~J., Lacey, C.~G., Baugh, C.~M., Cole, S., \& Frenk, C.~S.,  
2002, 
\mnras, 333, 156

\bibitem[Blackburne et al.(2006)]{blackburne} 
Blackburne, J.~A., Pooley, D., \& Rappaport, S.,  
2006,
\apj, 640, 569

\bibitem[Brada{\v c} et al.(2004)]{bradac2} 
Brada{\v c}, M., Schneider, P., Lombardi, M., Steinmetz, M., Koopmans, 
L.~V.~E., \& Navarro, J.~F.,  
2004, 
\aap, 423, 797

\bibitem[Brada{\v c} et al.(2002)]{bradac1} 
Brada{\v c}, M., Schneider, P., Steinmetz, M., Lombardi, M., King, L.~J., 
\& Porcas, R.,  
2002, 
\aap, 388, 373

\bibitem[Browne et al.(2003)]{browne}
Browne, I.~W.~A., et al.,  
2003,
\mnras, 341, 13

\bibitem[Bullock et al.(2000)]{bullock} 
Bullock, J.~S., Kravtsov, A.~V., \& Weinberg, D.~H.,  
2000,
\apj, 539, 517

\bibitem[Chartas et al.(2004)]{chartas}
Chartas, G., Dai, X., \& Garmire, G. P.,
2004, 
in Carnegie Astrophys. Ser. 2, Measuring and Modeling the Universe, 
ed. W. L. Freedman (Pasadena, CA: Carnegie Observatories);
http://www.ociw.edu/ociw/symposia/series /symposium2/proceedings.html

\bibitem[Chen et al.(2003)]{chenLOS} 
Chen, J., Kravtsov, A.~V., \& Keeton, C.~R.,  
2003, 
\apj, 592, 24

\bibitem[Chen et al.(2007)]{chenastro}
Chen, J., Rozo, E., Dalal, N., \& Taylor, J.~E.,
2007,
\apj, 659, 52

\bibitem[Cheng et al.(2002)]{cheng} 
Cheng, H.-C., Feng, J.~L., \& Matchev, K.~T.,  
2002, 
\prl, 89, 211301

\bibitem[Chiba(2002)]{chiba}
Chiba, M.,
2002,
\apj, 565, 17

\bibitem[Col{\'{\i}}n et al.(2000)]{WDM} 
Col{\'{\i}}n, P., Avila-Reese, V., \& Valenzuela, O.,  
2000, 
\apj, 542, 622

\bibitem[Colley et al.(2003)]{colley}
Colley, W. N., et al.,
2003,
\apj, 587, 71

\bibitem[Congdon \& Keeton(2005)]{congdon-mpole}
Congdon, A. B., \& Keeton, C. R.,
2005,
\mnras, 364, 1459

\bibitem[Congdon et al.(2007)]{congdon-micro} 
Congdon, A.~B., Keeton, C.~R., \& Osmer, S.~J.,  
2007, 
\mnras, 376, 263

\bibitem[Dalal \& Kochanek(2002)]{DK}
Dalal, N., \& Kochanek, C. S.,
2002,
\apj, 572, 25

\bibitem[Dav{\'e} et al.(2001)]{SIDM} 
Dav{\'e}, R., Spergel, D.~N., Steinhardt, P.~J., \& Wandelt, B.~D.,  
2001, 
\apj, 547, 574

\bibitem[Diemand et al.(2007)]{Diemand07}
Diemand, J., Kuhlen, M, \& Madau, P. 2007, \apj, 657, 262

\bibitem[Dobler \& Keeton(2006)]{dobler1}
Dobler, G., \& Keeton, C. R.,
2006,
\mnras, 365, 1243

\bibitem[Dodelson \& Widrow(1994)]{dodelson} 
Dodelson, S., \& Widrow, L.~M.,  
1994, 
\prl, 72, 17

\bibitem[Evans \& Witt(2003)]{EW}
Evans, N. W., \& Witt, H. J.,
2003,
\mnras, 345, 1351

\bibitem[Falco et al.(1985)]{FGS}
Falco, E. E., Gorenstein, M. V., \& Shapiro, I. I.,
1985,
\apjl, 289, L1

\bibitem[Fassnacht et al.(2004)]{LSSTlens} 
Fassnacht, C.~D., Marshall, P.~J., Baltz, A.~E., Blandford, R.~D., 
Schechter, P.~L., \& Tyson, J.~A.,  
2004,
BASS, 36, 1531

\bibitem[Feng(2005)]{feng} 
Feng, J.~L.,  
2005, 
Ann. Phys., 315, 2

\bibitem[Gao et al.(2004)]{Gao04}
Gao, L, White, S. D. M., Jenkins, A., Stoehr, F., \& Springel, V.,
2004, MNRAS, 355, 819 

\bibitem[Ghigna et al.(2000)]{Ghigna00}
Ghigna, S., Moore, F., Governato, F., Lake, G., Quinn, T., \& Stadel, J.,
2000, \apj, 544, 616

\bibitem[Helmi et al.(2002)]{Helmi02}
Helmi, A., White, S. D. M., \& Springel, V.,
2002, Phys. Rev. D, 666, 6, 063502

\bibitem[Heyl, Hernquist \& Spergel(1994)]{heyl}
Heyl, J. S., Hernquist, L., \& Spergel, D. N. 1994, \apj, 427, 165

\bibitem[Jesseit et al.(2005)]{JNB05}
Jesseit, R., Naab, T., \& Burkert, A.,  
2005,
\mnras, 360, 1185

\bibitem[Kundi{\'c} et al.(1997)]{kundic97}
Kundi{\'c}, T., Cohen, J. G., Blandford, R. D., \& Lubin, L. M.,
1997, \aj, 114, 507

\bibitem[Keeton(2001)]{gravlens}
Keeton, C. R., 2001, astro-ph/0102340;
http://redfive.rutgers.edu/ \~{}keeton/gravlens

\bibitem[Keeton(2003)]{anal}
Keeton, C. R.,
2003,
\apj, 584, 664

\bibitem[Keeton et al.(2006)]{0924}
Keeton, C. R., Burles, S., Schechter, P. L., \& Wambsganss, J.,
2006,
\apj, 639, 1

\bibitem[Keeton, Gaudi \& Petters(2003)]{cuspreln}
Keeton, C. R., Gaudi, B. S., \& Petters, A. O.,
2003,
\apj, 598, 138

\bibitem[Keeton, Gaudi \& Petters(2005)]{foldreln}
Keeton, C. R., Gaudi, B. S., \& Petters, A. O.,
2005,
\apj, 635, 35

\bibitem[Keeton \& Kochanek(1997)]{kk1115}
Keeton, C. R., \& Kochanek, C. S.,
1997,
\apj, 487, 42

\bibitem[Keeton \& Zabludoff(2004)]{KZ04}
Keeton, C. R., \& Zabludoff, A. I.,
2004,
\apj, 612, 660

\bibitem[Klypin et al.(1999)]{klypin}
Klypin, A., Kravtsov, A. V., Valenzuela, O., \& Prada, F.,
1999,
\apj, 522, 82

\bibitem[Kochanek(2002)]{cskH0}
Kochanek, C. S.,
2002,
\apj, 578, 25

\bibitem[Kochanek et al.(2007)]{csk-micro} 
Kochanek, C.~S., Dai, X., Morgan, C., Morgan, N., Poindexter, S., \& 
Chartas, G.,  
2007,
in ASP Conf. Ser. 43, Statistical Challenges in Modern Astronomy IV,
ed. G. J. Babu \& E. D. Feigelson (San Francisco, CA: ASP), 43;
arXiv:astro-ph/0609112

\bibitem[Kochanek \& Dalal(2004)]{KD}
Kochanek, C.~S., \& Dalal, N.,  
2004,
\apj, 610, 69

\bibitem[Kochanek et al.(2006)]{csk-find} 
Kochanek, C.~S., Mochejska, B., Morgan, N.~D., \& Stanek, K.~Z.,  
2006,
\apjl, 637, L73

\bibitem[Kochanek et al.(2004)]{saasfee}
Kochanek, C. S., Schneider, P., \& Wambsganss, J., 2004, Part 2 of
Gravitational Lensing: Strong, Weak \& Micro, Proceedings of the 33rd
Saas-Fee Advanced Course, ed. G. Meylan, P. Jetzer, \& P. North
(Berlin: Springer-Verlag);
arXiv:astro-ph/0407232

\bibitem[Koopmans et al.(2004)]{SKAlens} 
Koopmans, L.~V.~E., Browne, I.~W.~A., \& Jackson, N.~J.,  
2004, 
New Astron. Rev., 48, 1085

\bibitem[Koopmans \& de Bruyn(2000)]{1600micro} 
Koopmans, L.~V.~E., \& de Bruyn, A.~G.,  
2000, 
\aap, 358, 793

\bibitem[Koposov et al.(2008)]{koposov} 
Koposov, S., et al.,  
2008,
\apj, 686, 279

\bibitem[Koposov et al.(2009)]{koposov2}
Koposov, S. E., Yoo, J., Rix, H.-W., Weinberg, D. H., Macci{\`o}, A. V.,
\& Miralda-Escud{\'e}, J.,
2009,
\apj, 696, 2179

\bibitem[Koushiappas et al.(2004)]{koush} 
Koushiappas, S.~M., Zentner, A.~R., \& Walker, T.~P.,  
2004, 
\prd, 69, 043501

\bibitem[Kravtsov et al.(2004)]{kravtsov} 
Kravtsov, A.~V., Gnedin, O.~Y., \& Klypin, A.~A.,  
2004,
\apj, 609, 482

\bibitem[Kuhlen et al.(2004)]{kuhlen} 
Kuhlen, M., Keeton, C.~R., \& Madau, P.,  
2004,
\apj, 601, 104

\bibitem[Lieu(2008)]{lieu} 
Lieu, R.,  
2008,
\apj, 674, 75

\bibitem[Macci{\`o} et al.(2009)]{maccio2} 
Macci{\`o}, A.~V., Kang, X., Fontanot, F., Somerville, R. S.,
Koposov, S., \& Monaco, P.
2009, 
arXiv:0903.4681

\bibitem[Macci{\`o} et al.(2006)]{maccio} 
Macci{\`o}, A.~V., Moore, B., Stadel, J., \& Diemand, J.,  
2006, 
\mnras, 366, 1529

\bibitem[Mao \& Schneider(1998)]{MS}
Mao, S., \& Schneider, P.,
1998,
\mnras, 295, 587

\bibitem[Marshall et al.(2005)]{SNAPlens} 
Marshall, P., Blandford, R., \& Sako, M.,  
2005,
New Astron. Rev., 49, 387

\bibitem[Metcalf(2005a)]{metcalfLOS1} 
Metcalf, R.~B.,  
2005a, 
\apj, 622, 72

\bibitem[Metcalf(2005b)]{metcalfLOS2} 
Metcalf, R.~B.,  
2005b, 
\apj, 629, 673

\bibitem[Metcalf \& Madau(2001)]{MM}
Metcalf, R. B., \& Madau, P.,
2001,
\apj, 563, 9

\bibitem[Metcalf et al.(2004)]{metcalf2237} 
Metcalf, R.~B., Moustakas, L.~A., Bunker, A.~J., \& Parry, I.~R.,  
2004, 
\apj, 607, 43

\bibitem[Metcalf \& Zhao(2002)]{MZ02}
Metcalf, R. B., \& Zhao, H.,
2002
\apj, 567, L5

\bibitem[Momcheva et al.(2006)]{iva}
Momcheva, I., Williams, K., Keeton, C., \& Zabludoff, A.,
2006, \apj, 641, 169

\bibitem[Moore et al.(1999)]{moore}
Moore, B., Ghigna, S., Governato, F., Lake, G., Quinn, T., Stadel, J.,
\&  Tozzi, P.,
1999,
\apjl, 524, L19

\bibitem[Morgan et al.(2006)]{1131tdel}
Morgan, N. D., Kochanek, C. S., Falco, E. E., \& Dai, X.,
2006,
arXiv:astro-ph/0605321

\bibitem[Moustakas et al.(2008)]{omega}
Moustakas, L. A., et al.
2008,
in Proc. SPIE 7010, Space Telescopes and Instrumentation 2008: Optical,
Infrared, and Millimeter, 
ed. J. M. Oschmann, M. W. M. de Graauw \& H. A. MacEwen;
arXiv:0806.1884

\bibitem[Naab \& Burkert(2003)]{NB03}
Naab, T., \& Burkert, A.,  
2003,
\apj, 597, 893

\bibitem[Naab et al.(2006)]{naab06}
Naab, T., Jesseit, R., \& Burkert, A.,  
2006,
\mnras, 372, 839

\bibitem[Oguri(2007)]{oguriH0}
Oguri, M.,
2007,
\apj, 660, 1

\bibitem[Oguri \& Lee(2004)]{oguri-sub} 
Oguri, M., \& Lee, J.,  
2004, 
\mnras, 355, 120

\bibitem[Pooley et al.(2007)]{pooley} 
Pooley, D., Blackburne, J.~A., Rappaport, S., \& Schechter, P.~L.,  
2007,
\apj, 661, 19

\bibitem[Rest et al.(2001)]{rest}
Rest, A., van den Bosch, F. C., Jaffe, W., Tran, H., Tsvetanov, Z.,
Ford, H. C., Davies, J., \& Schafer, J. 2001, \aj, 121, 2431

\bibitem[Rix et al.(2004)]{GEMS}
Rix, H.-W., et al.,
2004,
\apjs, 152, 163


\bibitem[Saglia, Bender \& Dressler(1993)]{saglia}
Saglia, R. P., Bender, R., \& Dressler, A. 1993, \aap, 279, 75

\bibitem[Saha(2000)]{saha-degen}
Saha, P.,
2000,
\aj, 120, 1654

\bibitem[Saha \& Williams(2003)]{SW-tdel}
Saha, P., \& Williams, L. L. R.,
2003,
\aj, 125, 2769


\bibitem[Schechter \& Wambsganss(2002)]{SW02}
Schechter, P. L., \& Wambsganss, J.,
2002,
\apj, 580, 685

\bibitem[Schneider, Ehlers \& Falco(1992)]{SEF}
Schneider, P., Ehlers, J., \& Falco, E. E.,
1992,
Gravitational Lenses (Berlin: Springer-Verlag)

\bibitem[Shin \& Evans(2008)]{shin}
Shin, E. M., \& Evans, N. W.,
2008,
\mnras, 385, 2107

\bibitem[Sluse et al.(2003)]{sluse}
Sluse, D., et al., 2003,
\aap, 406, L43

\bibitem[Somerville(2002)]{somerville} 
Somerville, R.~S.,  
2002,
\apjl, 572, L23

\bibitem[Strigari et al.(2007a)]{missing} 
Strigari, L.~E., Bullock, J.~S., Kaplinghat, M., Diemand, J., Kuhlen, M., 
\& Madau, P.,  
2007a, 
\apj, 669, 676

\bibitem[Strigari et al.(2007b)]{strigari-decay} 
Strigari, L.~E., Kaplinghat, M., \& Bullock, J.~S.,  
2007b, 
\prd, 75, 061303

\bibitem[Taylor \& Babul(2001)]{TB01} 
Taylor, J.~E., \& Babul, A.,  
2001, 
\apj, 559, 716

\bibitem[Taylor \& Babul(2004)]{TB04} 
Taylor, J.~E., \& Babul, A.,  
2004, 
\mnras, 348, 811

\bibitem[van den Bosch et al.(2005)]{vdb} 
van den Bosch, F.~C., Tormen, G., \& Giocoli, C.,  
2005, 
\mnras, 359, 1029

\bibitem[Weymann et al.(1980)]{weymann}
Weymann, R.~J., et al.,  1980,
\nat, 285, 641

\bibitem[Witt et al.(1995)]{WMS95} 
Witt, H.~J., Mao, S., \& Schechter, P.~L.,  
1995, 
\apj, 443, 18

\bibitem[Wyithe et al.(2000)]{wyithe} 
Wyithe, J.~S.~B., Webster, R.~L., Turner, E.~L., \& Mortlock, D.~J.,  
2000,
\mnras, 315, 62

\bibitem[Yoo et al.(2005)]{yoo1}
Yoo, J., Kochanek, C. S., Falco, E. E., \& McLeod, B. A.,
2005, \apj, 626, 51

\bibitem[Yoo et al.(2006)]{yoo2}
Yoo, J., Kochanek, C. S., Falco, E. E., \& McLeod, B. A.,
2006, \apj, 642, 22

\bibitem[Zentner et al.(2005)]{Z05} 
Zentner, A.~R., Berlind, A.~A., Bullock, J.~S., Kravtsov, A.~V., \& 
Wechsler, R.~H.,  
2005, 
\apj, 624, 505

\bibitem[Zentner \& Bullock(2003)]{ZB03} 
Zentner, A.~R., \& Bullock, J.~S.,  
2003, 
\apj, 598, 49

\end{thebibliography}
\end{document}